\documentclass[useAMS,usenatbib]{mn2e}

\usepackage{defs}
\usepackage{epsfig}
\usepackage{graphicx}
\usepackage{amsmath}
\usepackage{amssymb}
\usepackage{lscape}
\usepackage{url}
\usepackage{setspace}
\usepackage{float}

\title[PAH Emission in Powerful HzRGs]{PAH Emission in Powerful High-Redshift Radio Galaxies} 
\author[J. I. Rawlings et al.]
{\parbox{\textwidth}{\raggedright J.I.~Rawlings$^{1}$\thanks{E-mail: \texttt{jir2@mssl.ucl.ac.uk}},
N.~Seymour$^{2,1}$, M.J.~Page$^{1}$, C.~De Breuck$^{3}$, D.~Stern$^{4}$, M.~Symeonidis$^{1}$, P.N.~Appleton$^{5}$, A.~Dey$^{6}$, M.~Dickinson$^{6}$, M.~Huynh$^{7}$, E.~Le Floc'h$^{8}$, M.~Lehnert$^{9}$, J.R.~Mullaney$^{8,10}$, N.~Nesvadba$^{9}$, P.~Ogle$^{7}$, A.~Sajina$^{11,12}$, J.~Vernet$^{3}$ and A.~Zirm$^{13}$}\vspace{0.4cm}\\
$^{1}$ Mullard Space Science Laboratory, University College London, Holmbury St. Mary, Dorking, Surrey, RH5 6NT, UK\\
$^{2}$ CSIRO Astronomy \& Space Science, PO Box 76, Epping, NSW, 1710, Australia\\
$^{3}$ European Southern Observatory, Karl Schwarzschild Strae 2, 85748 Garching, Germany\\
$^{4}$ Jet Propulsion Laboratory, California Institute of Technology, Pasadena, CA 91109, USA\\
$^{5}$ HSC, California Institute of Technology, Mail Code 220-6, Pasadena, CA 91125, USA\\
$^{6}$ NOAO-Tucson, 950 North Cherry Avenue, Tucson, AZ 85719, USA\\
$^{7}$ IPAC, California Institute of Technology, Mail Code 220-6, Pasadena, CA 91125, USA\\ 
$^{8}$ Laboratoire AIM, CEA/DSM/IRFU, CNRS, Universit\'é  Paris-Diderot, 91191 Gif-sur-Yvette Cedex, France\\
$^{9}$ Leiden Observatory, University of Leiden, P.O. Box 9513, 2300 RA Leiden, Netherlands\\
$^{10}$ Department of Physics, Durham University, South Road, Durham DH1 3LE 2, UK\\
$^{11}$ Department of Physics and Astronomy, Tufts University, Medford, MA 02155, USA\\
$^{12}$ Department of Physics and Astronomy, Haverford College, Haverford, PA 19041, USA\\
$^{13}$ Dark Cosmology Centre, Niels Bohr Institute, University of Copenhagen, Juliane Maries Vej 30, Denmark}

\date{Accepted 2012 November 6.  Received 2012 November 6; in original form 2012 July 18}
\pagerange{\pageref{firstpage}--\pageref{lastpage}}
\pubyear{2012}

\begin{document}

\maketitle

\label{firstpage}

\begin{abstract}
We present the mid-infrared spectra of seven of the most powerful radio-galaxies known to exist at $1.5\,<\,z\,<\,2.6$. The radio emission of these sources is dominated by the AGN with 500\,MHz luminosities in the range 10$^{27.8}-$10$^{29.1}$\,W\,Hz$^{-1}$. The AGN signature is clearly evident in the mid-infrared spectra, however, we also detect polycyclic aromatic hydrocarbons emission, indicative of prodigious star formation at a rate of up to $\sim$\,1000\,M$_{\rm \odot}$\,yr$^{-1}$. Interestingly, we observe no significant correlation between AGN power and star formation in the host galaxy. We also find most of these radio galaxies to have weak 9.7\,$\mu$m silicate absorption features ($\tau_{\rm 9.7\,\mu m}\,<\,0.8$) which implies that their mid-IR obscuration is predominantly due to the dusty torus that surrounds the central engine, rather than the host galaxy. The tori are likely to have an inhomogeneous distribution with the obscuring structure consisting of individual clouds. We estimate that these radio galaxies have already formed the bulk of their stellar mass and appear to lie at a stage in their evolution where the obscured AGN dominates the energy output of the system but star formation is also prevalent.
\end{abstract}


\section{Introduction}
It is now understood that supermassive black holes (SMBHs) are at the centres of all massive galaxies \citep{Magorrian:98, Richstone:98}. These SMBHs can be actively accreting \citep{Blandford:82, Churazov:05, Russell:10} which has given rise to the population of active galactic nuclei (AGN). A small fraction \citep[$<$\,10\%,][]{Tasse:08} of these AGN are `radio-loud' in that their radio flux is greater than their \textit{R}-band flux by a factor $>$\,10. More specifically however, the radio-loud fraction may vary with redshift, mass and optical luminosity \citep{Jiang:07}. Powerful radio-loud AGN show the presence of twin jets of relativistic particles which emit synchrotron radiation at radio wavelengths \citep{Lynden-Bell:69}. The radio jets can reach up to hundreds of kilo-parsecs in length and extend well beyond the realms of the host galaxy. Under the unification model \citep{Rowan-Robinson:77, Lawrence:87}, radio galaxies are the hosts of radio-loud type-2 AGN, whose jets are oriented approximately perpendicular to our line of sight and much of the central engine is obscured by a dusty torus \citep{Antonucci:84, Barthel:89}, allowing the host galaxy to outshine the AGN in the optical.

In the local Universe, radio galaxies are associated with massive giant elliptical galaxies \citep{Matthews:64}. Studies of radio galaxies at high redshift \citep{Pentericci:01, Zirm:03, Seymour:07, DeBreuck:10} indicate that they too are massive systems (10$^{10}$-10$^{12}$\,M$_{\rm \odot}$) up to at least $z\,\sim$\,5.2. As a result, it has been suggested that high-redshift radio galaxies (HzRGs) are the precursors of the modern day elliptical galaxies that reside in rich cluster environments \citep{Archibald:01, Stevens:03, Reuland:04, Miley:08, Galametz:12, Mayo:12}. HzRGs have been found to host SMBHs of 10$^{9}$-10$^{10}$\,M$_{\rm \odot}$ \citep{Nesvadba:11}. They also have higher accretion rates than their local counterparts as inferred from mid-infrared (mid-IR) continuum luminosities (e.g. rest-frame $\nu L_{\rm \nu}$\,(15\,$\mu$m)\,$>$\,10$^{44}$\,erg\,s$^{-1}$ for $z\,>$\,0.5, \citealt{Ogle:06}; see also \citealt{Seymour:07, Sajina:07a, DeBreuck:10, Ogle:07, Ogle:10}).

From other investigations into the host properties of HzRGs, there is tentative evidence that some have very high star formation rates (SFRs). For example, from sub-mm observations of radio galaxies \citep{Dunlop:94, Archibald:01, Reuland:04, Greve:06}, SFRs of $\sim$\,1000\,M$_{\rm \odot}$\,yr$^{-1}$ have been reported, while SFRs of up to 1500\,M$_{\rm \odot}$\,yr$^{-1}$ were derived from the luminous ($>$\,100\,kpc) Ly$\alpha$ halos around some radio galaxies \citep{Reuland:03, VillarMartin:03}. Furthermore, \citet{Seymour:11} investigated the star-forming properties of radio-loud AGN in the far-IR and while they made no distinction between type-1 and type-2 AGN, for their high-redshift ($1.2\,<\,z\,<\,3$) sample they found a mean SFR range of 80-580\,M$_{\rm \odot}$\,yr$^{-1}$. 

The spectral energy distributions (SEDs) of radio galaxies show copious amounts of emission from both the AGN and host galaxy. In the mid-IR, along with reprocessed emission from dust associated with the AGN torus \citep{Neugebauer:87}, polycyclic aromatic hydrocarbon (PAH) emission features \citep{Puget:89, Allamandola:99} may be present. PAH emission manifests itself as broad ($\sim$\,1\,$\mu$m) features, primarily at 6.2, 7.7, 8.6 and 11.3\,$\mu$m, that predominantly appear in the mid-IR spectra of dusty galaxies \citep{Helou:00, Smith:07}. The various studies on AGN-starburst composite sources show that PAHs can survive in galaxies that contain an active nucleus \citep{Desai:07, Armus:07, Treyer:10}. PAH emission can be used as an approximate measure of a galaxy's SFR as it is thought that PAHs become excited by the absorption of UV emission from massive, young, hot stars. The relation is further supported by the observed correlation between PAHs and the cold dust reservoirs associated with star formation \citep[hereafter P08]{Haas:02, Peeters:04, Lacy:07, Sajina:08, Pope:08}. It has been suggested that PAHs could be present in local radio galaxies \citep{Siebenmorgen:04}. However, their weak detection or absence in other studies \citep{Haas:05, Ogle:07, Ogle:10, Nesvadba:10} suggest considerably lower SFRs ($<$\,5\,M$_{\odot}$\,yr$^{-1}$) than those detected in radio-loud AGN at high-redshift \citep{Sajina:07b, Seymour:08, Seymour:11, Ogle:12}. From the few mid-IR spectra of HzRGs to date, PAH emission has been detected on a couple of occasions \citep{Seymour:08, Ogle:12}. This has given SFRs of up to a few 1000\,M$_{\odot}$\,yr$^{-1}$ which compliment the sub-mm based estimates. 

Another feature that may be present in the mid-IR spectra of radio galaxies is silicate absorption at 9.7\,$\mu$m and 18\,$\mu$m \citep{Genzel:00, Yan:07, Sajina:07a, Nikutta:09, Sales:11}. From the prevalence of weak silicate absorption features observed in type-2 AGN, it has been suggested that their obscuring tori are characterised by a clumpy, rather than smooth, distribution of dust \citep{Levenson:07, Spoon:07, Imanishi:07, Sajina:07a}. Strong absorption features on the other hand, have been interpreted as evidence for a large contribution from the host galaxy to the obscuration of the nucleus \citep{Ogle:06}. The weak silicate absorption feature at 9.7\,$\mu$m for one HzRG; 4C 23.56 \citep[$\tau_{\rm 9.7\,\mu m}$\,=\,0.3\,$\pm$\,0.05]{Seymour:08} is comparable to those found for some lower-redshift and less-luminous radio galaxies \citep{Ogle:06, Cleary:07}. However, a small number of sources from \citet{Ogle:06} showed deeper silicate absorption as did the high-redshift radio-loud AGN from \citet[$\tau_{\rm 9.7\,\mu m}\,>$\,1]{Sajina:07b}. However, the latter sample was selected to contain sources with significant obscuration. 

In this paper, we present the mid-IR spectra (5\,$\mu$m\,$<$\,$\lambda_{\rm rest}$\,$<$\,13\,$\mu$m) of seven HzRGs, the largest sample to date, that were obtained using the Infrared Spectrograph \citep[IRS,][]{Houck:04} instrument on-board the \textit{Spitzer Space Telescope} \citep{Werner:04}. We investigate the connection between star formation and the radio-loud phase of powerful obscured AGN by decomposing the spectra into AGN and host galaxy components. This allows the relative AGN and star-forming contributions to the mid-IR output to be determined (\citealt{Sajina:07a}; P08; \citealt{Menendez-Delmestre:09}, hereafter MD09; \citealt{Coppin:10, Petric:11, Fiolet:10, Mullaney:11}).

The paper is structured as follows: In \S\ref{Section:sample observations and reduction} we describe the sample, the observations and our data reduction pipeline, in \S\ref{Section:mid-IR spectral decomposition} we present our spectral fitting model, while in \S\ref{Section:results} and \S\ref{Section:discussion} we comment on our results and draw conclusions.

Throughout, we assume a $\Lambda$-dominated flat Universe, with H$_{\rm 0}$\,=\,70\,km\,s$^{-1}$\,Mpc$^{-1}$, $\Omega_{\rm \Lambda}$\,=\,0.7 and $\Omega_{\rm m}$\,=\,0.3. Unless otherwise stated, uncertainties are 1\,$\sigma$ and upper limits are 3\,$\sigma$.

\section{The sample, observations and reduction}
\label{Section:sample observations and reduction}
\subsection{Sample selection}
\label{Section:sample selection}
Our sample of radio galaxies originate from the \textit{Spitzer} High-redshift Radio Galaxy (SHzRG) project \citep[hereafter DB10]{Seymour:07, DeBreuck:10} which focused on targeting the host galaxy properties of some of the most powerful obscured radio-loud AGN in the Universe. The SHzRG sources were selected from multiple radio galaxy surveys \citep[for list and references, see][]{Seymour:07} for which radio flux densities were obtained from all-sky, low-frequency (74, 327/352, 365 and 1400 MHz) surveys \citep{Rengelink:97, Douglas:96, Condon:98, Cohen:07}. They were selected to uniformly sample the redshift-radio luminosity plane for $z>1$ and $L_{\rm 3\,GHz}>10^{26}$\,W\,Hz$^{-1}$. Radio galaxies with ancillary data were preferred. With these selection criteria, this gave a sample of 70 objects that is representative of the overall HzRG population. Spectroscopic redshifts and rest-frame 3\,GHz luminosities were in the range $1<z<5.2$ and $L_{\rm 3\,GHz}$\,=\,10$^{26.2}$-10$^{28.7}$\,W\,Hz$^{-1}$, respectively. Due to their extremely high radio luminosities, objects of this nature are very rare, with only a few hundred known at $z>1$ across the whole sky. Most have steep spectra in the radio \citep{Ker:12} with their 3\,GHz synchrotron emission free of Doppler-boosting \citep{Seymour:07}. Rest-frame 500\,MHz luminosities, $L_{\rm 500\,MHz}$, were calculated by interpolating between the luminosities from the different surveys. The isotropy of the synchrotron emission is expected to anti-correlate with frequency \citep{Blundell:98} and 500\,MHz was the lowest rest-frequency, for these redshifts, at which luminosities could be interpolated. The SHzRG objects were first observed with all seven photometric bands of the \textit{Spitzer}-Infrared Array Camera \citep[IRAC,][]{Fazio:04} and Multiband Imaging Photometer for \textit{Spitzer} \citep[MIPS,][]{Rieke:04} instruments. Later, for a select few, we used the IRS for follow-up spectroscopy. 

\begin{figure*}
\begin{tabular}{c}
\epsfig{file=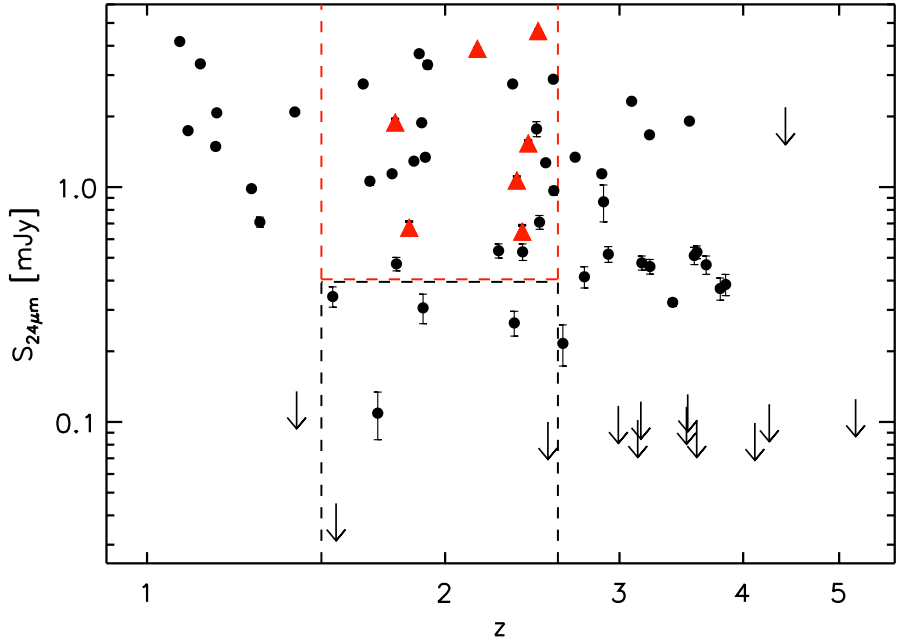,width=0.75\linewidth}\\
\end{tabular}
\caption{Observed 24\,$\mu$m flux density as a function of redshift for the whole SHzRG sample. The radio galaxies in our sample are represented by red triangles. Sources within the region enclosed by the red dashed lines correspond to SHzRG sources that are within our redshift range and `bright' at 24\,$\mu$m (S$_{\rm 24\mu m}$\,$>$\,0.4\,mJy). Sources within the region enclosed by the black dashed lines correspond to SHzRG sources that are within our redshift range and `faint' at 24\,$\mu$m (S$_{\rm 24\mu m}$\,$<$\,0.4\,mJy).}
\label{Figure:sample}
\end{figure*}

The sample presented in this paper consists of seven of the eight SHzRG objects that have been observed with the IRS and lie in the redshift range $1.5\,<\,z\,<\,2.6$. We reject the radio galaxy PKS 0529-549 ($z$\,=\,2.575) as there is a source present in the 24\,$\mu$m MIPS map that cannot be distinguished from the target in the 2-d IRS data. Since the unknown source lies within the target extraction window, this makes the IRS spectrum of PKS 0529-549 unreliable. Our sample includes the IRS observation of 4C 23.56 previously presented by \citet{Seymour:08}, though we use a more rigorous spectral reduction pipeline here. We do not include TXJ 1908+7220 in the sample as its higher-redshift ($z$\,=\,3.53) results in the 9.7\,$\mu$m silicate feature and a significant fraction of the 7.7\,$\mu$m PAH feature being shifted outside of the IRS spectral coverage. The properties of our sample are given in Table \ref{Table:source_info} and we show the 24\,$\mu$m fluxes as a function of redshift for the whole SHzRG sample in Figure \ref{Figure:sample}. The IRS observations are from multiple proposals, with each proposal having its own particular science goals and we can treat the selection of our sample as being a random selection from mid-IR `bright' radio galaxies. All targets are mid-IR `bright' in that they have 24\,$\mu$m flux densities, S$_{\rm 24\mu m}$\,$>$\,0.4\,mJy which ensured that good quality spectra (S/N\,$\gtrsim$\,5 per resolution element) could be produced in reasonable observing times. More sources from some of these proposals would have been observed with the IRS, if it was not for the depletion of the on-board cryogen. Four radio galaxies in our sample (3C 239, 4C +23.56, NVSS J171414$+$501530 and PKS 1138$-$26) have also been observed at sub-mm wavelengths with the Submillimetre Common-User Bolometer Array (SCUBA) on the James Clerk Maxwell Telescope (JCMT). Their sub-mm emission is discussed in \S\ref{Section:discussion}.

\subsection{Observations}
The targets were observed with the Long-Low (LL) module (resolving power, R\,$\sim$\,60-130) of the IRS. While all targets were observed in the first order of the LL module (LL1; 20\,$\mu$m\,$\lesssim$\,$\lambda$\,$\lesssim$\,40\,$\mu$m), only three were also observed in the second order (LL2; 14\,$\mu$m\,$\lesssim$\,$\lambda$\,$\lesssim$\,21\,$\mu$m). The use of the LL2 order ensured the spectra of the objects at lower redshift covered the shorter end of our required rest-frame wavelength range of 6-10\,$\mu$m. All but PKS 1138$-$26 were observed in staring-mode. In staring-mode, the IRS observes a target in two positions (`nods') in the slit of each order, resulting in four different 2-d spectral images if a target is observed in both the LL1 and LL2 orders (two images if the target is just observed in the LL1 order). Observations are then repeated in cycles for each slit and position. It was anticipated that the required S/N could be obtained for each spectrum by running observations of 20 cycles or fewer of 120\,s duration for each target. 

PKS 1138$-$26 was observed in mapping-mode, where the target was observed at eight positions, in 1.67-arcsec steps, along the LL1 slit. This single astronomical observation request (AOR) was coupled with eight other AORs to spectrally map the environment of PKS 1138$-$26 as part of a larger observation. The sampling of these eight AORs was different from the single AOR and it was intended that a separate, higher S/N spectrum be produced alongside the spectrum from the single AOR. \citet{Ogle:12} describes these AORs in more detail and presents the higher S/N spectrum of PKS 1138$-$26.

For each target, the number of cycles varied depending on the exposure times required to reach the S/N threshold. The details of the observations are given in Table \ref{Table:source_obs}.

\subsection{Reduction pipeline}
\label{Section:reduction pipeline}
For the seven targets observed in staring-mode, the LL module data obtained from the \textit{Spitzer} Science Center (SSC) came in the form of Basic Calibrated Data (BCD) images produced by the S18.18 version of the IRS calibration software pipeline. Included in the pipeline were the low-level processes: ramp fitting, dark sky subtraction, linearity correction, flat-fielding, droop correction, straylight/cross-talk correction, wavelength calibration and flux conversion. In addition to the SSC pipeline processing, bad pixel detection and cleaning, latent charge removal and background subtraction were performed using our own reduction pipeline. The steps of this pipeline closely followed those outlined in \citet{Goulding_thesis}. The pipeline included masking damaged or unreliable pixels, either those previously flagged by the SSC calibration files or those deemed to be too variant (5\,$\sigma$) from their neighbouring pixels or their counterparts in other images. Masking such pixels meant they would not be considered when combining the images from individual observations. Images for each `nod' in each order were combined using a clipped median (5\,$\sigma$). Also performed on individual images was the subtraction of the latent charge that can remain on the detector between frames \citep{Teplitz:07}.

After the 2-d images for each `nod' in each order were combined, the IRSCLEAN tool was used to replace any masked pixel values by interpolating the surrounding pixel values, effectively `cleaning' the images. The next step was to subtract the background. For each combined image, a corresponding background image was created from as many of the other combined images for that source as possible. An important factor in the number of combined images used to create the `background' image was the presence of serendipitous sources. A combined image would not be used for the background if it was deemed to have a serendipitous source near the target-observing position as this would result in a contaminated background for the target. IRSCLEAN was then run again, on the background-subtracted images.

The final step in the pipeline was to perform the spectral extraction on the 2-d background-subtracted images. For this we used the Spectroscopy Modeling Analysis Reduction Tool (SMART) \citep{Higdon:04}. At the spatial resolution of \textit{Spitzer}, our targets are point sources, so we used the optimal extraction option inside SMART and set the extraction window to be centred on the target position. This produced a 1-d spectrum for each nod in each order. Flux uncertainties were then estimated from the fluctuations along each row (i.e. wavelength) of the 2-d images. The extracted 1-d spectra were combined outside SMART and the uncertainties combined in quadrature. The spectra can then be compared to the 16\,$\mu$m IRS blue peak-up and 24\,$\mu$m MIPS photometry.

Regarding PKS 1138$-$26, due to the finer sampling, the background could not be estimated from the single AOR. Instead a median background image was created from one of the eight AORs obtained by \citet{Ogle:12}. The images for each position were combined using the standard \textit{Spitzer} coadd pipeline and the background image was subtracted. The eight resulting images were then run through IRSCLEAN and the 1-d spectra were optimally extracted using the \textit{Spitzer} IRS Custom Extraction (SPICE) software. The final spectrum of PKS 1138$-$26 is the median of the 1-d spectra. The individual spectra are shown in Figure \ref{Figure:spectra}.

\subsection{The mean spectrum}
To obtain a measure of the average star-forming properties of our sample, we also create an un-weighted, arithmetic mean luminosity spectrum from the rest-frame spectra of the seven targets. So that the spectra have a constant weighting to the mean as a function of wavelength, we restrict the mean spectrum to only cover a wavelength range that all seven constituent spectra cover. This results in a coverage of 6.5-10\,$\mu$m. So that an accurate mean could be calculated, the original spectra are re-sampled. The re-sampling process is such that all seven spectra contribute to the mean in each resolution element and there is little degradation in the spectral resolution.

\section{Mid-IR spectral decomposition}
\label{Section:mid-IR spectral decomposition}
\subsection{The model and fitting process}
\label{Section:the model}
To decompose the spectra into contributions from AGN and star formation, we perform spectral fitting on the mean and individual spectra in rest-frame luminosity space using a variant\footnote{Our variant is that the procedure only involved IRS spectroscopy; i.e., no broad-band IR photometry was fitted.} of the {\tt DecompIR}\footnote{https://sites.google.com/site/decompir} IDL fitting procedure described in \citet{Mullaney:11}. For the host galaxy component we use a library of templates that comprise the IRS spectra of the \citet{Brandl:06} sample of local starburst galaxies. These templates have been re-reduced and are presented in \citet{Mullaney:11}. From the \citet{Brandl:06} sample, we exclude the galaxies that contain an AGN (Seyfert galaxies, low-ionisation nuclear emission-line region; LINERs, etc.) and accept only the pure starburst objects. This gives us 10 templates that differ in the strength of their PAH features and also their silicate absorption at 9.7\,$\mu$m. The AGN component is characterised by a power-law with a spectral index, $\alpha$, that is allowed to vary between 1.0 and 1.5. We place this restriction as similar ranges for the mid-IR spectral indices of type-1 AGN have been found (e.g. broad-line radio galaxies, \citealt{Ogle:06}; Seyfert galaxies, \citealt{Buchanan:06}; high-redshift ULIRGs, \citealt{Alonso-herrero:06} and type-1 local AGN, \citealt{Mullaney:11}) and the mid-IR slopes (5\,$\mu$m\,$<$\,$\lambda$\,$<$\,38\,$\mu$m) of type-1 and type-2 objects have been shown to be comparable \citep{Nenkova:08a}. To account for the higher inclinations of our type-2 sources, we make an assumption that the dusty environments of high-redshift radio-loud AGN can be modelled by a Galactic extinction curve \citep{Sajina:07a, Lacy:07, Seymour:08} which we apply to the power-law slope. We choose the \citet{Draine:03} extinction curve that uses R$_{\rm V}$\,=\,3.1, the typical value for the Milky Way Galaxy, where R$_{\rm V}$ is the ratio of total extinction, A(V), to selective extinction, A(B)-A(V), in the V band (i.e. R$_{\rm V}$\,=\,A(V)/(A(B)-A(V))). Although this is a simplistic approach as we assume that the obscuring medium acts as a simple foreground screen, it will suffice for fitting AGN continua and silicate absorption features in our spectra. We do not apply extinction to the host galaxy component as the \citet{Brandl:06} spectra are empirical and have a certain level of obscuration already, evident through their silicate absorption features. 

The five free parameters for our model are: (i) the host galaxy template, (ii) the host galaxy normalisation, (iii) the power-law normalisation, (iv) power-law spectral index, $\alpha$ and (v) the hydrogen column density, $N_{\rm H}$. The expression for the modelled mid-IR spectrum, $L_{\rm \nu}$, in W\,Hz$^{-1}$ is thus:
\begin{equation}
{L_{\rm \nu}(\lambda)\,=\,k_{\rm 1}\,\lambda^{\alpha}\,e^{-N_{\rm H}\,\sigma(\lambda)}\,+\,k_{\rm 2}\,f_{\rm \nu}(\lambda)}
\end{equation}
Where $f_{\rm \nu}(\lambda)$ is the flux density, in Jy, of the host galaxy template, $k_{\rm 1}$ and $k_{\rm 2}$ are the normalisation factors, and $\sigma(\lambda)$ is the extinction cross-section that, with the hydrogen column density, gives the optical depth, $\tau(\lambda)$; i.e. $\tau(\lambda)$\,=\,$N_{\rm H}\,\sigma(\lambda)$. 

\begin{figure*}
\begin{tabular}{cc}
\epsfig{file=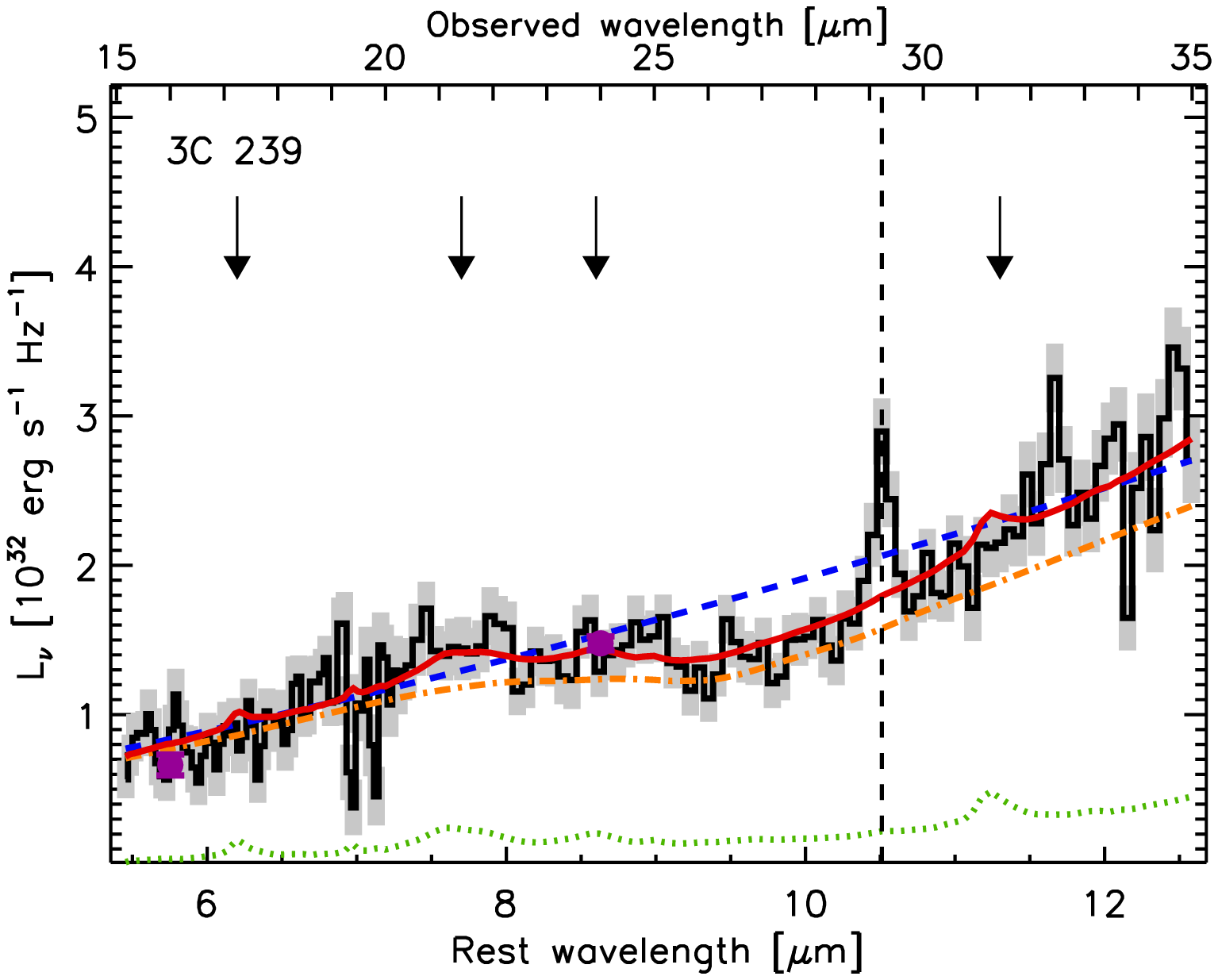,width=0.42\linewidth} &\epsfig{file=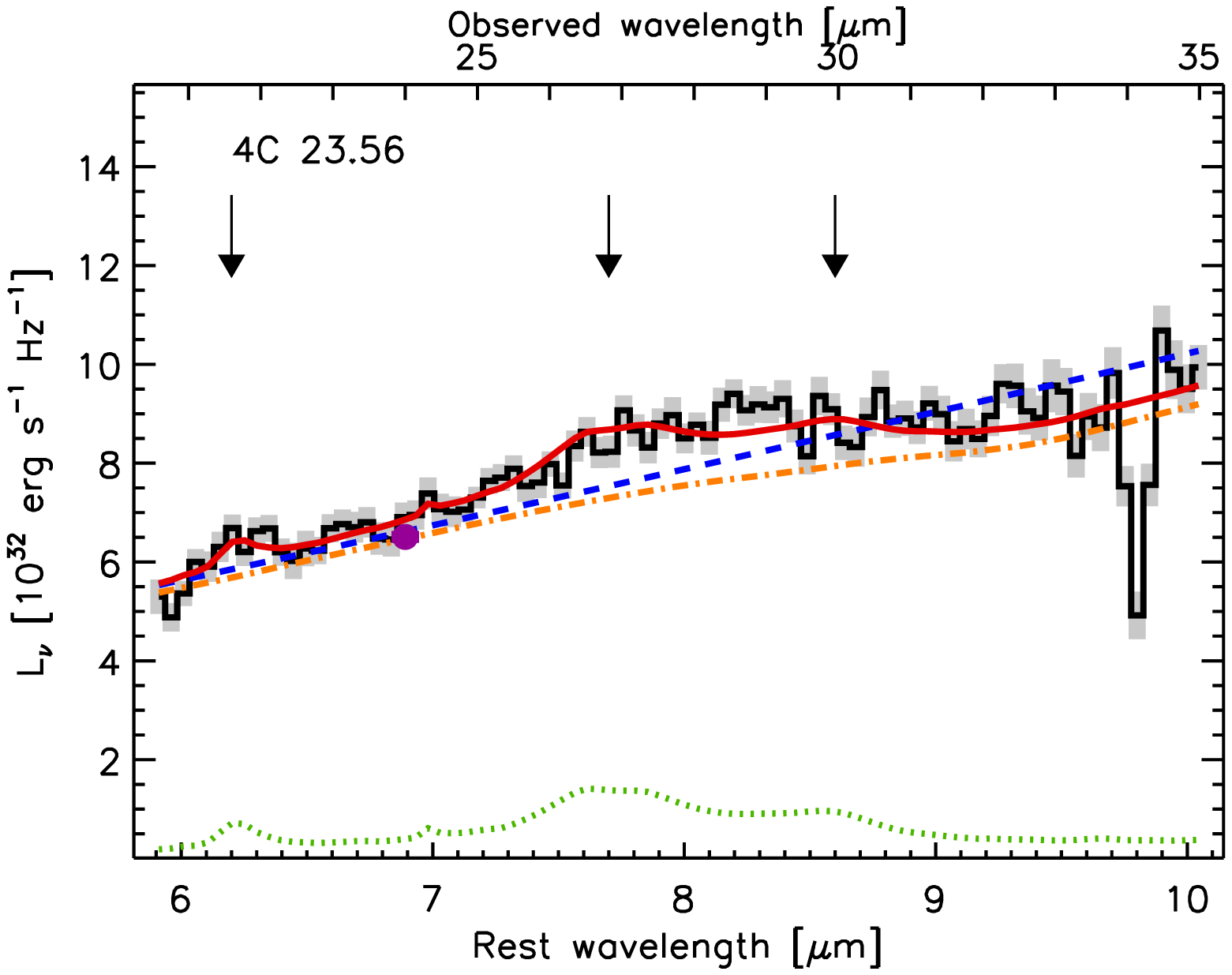,width=0.42\linewidth}\\
\epsfig{file=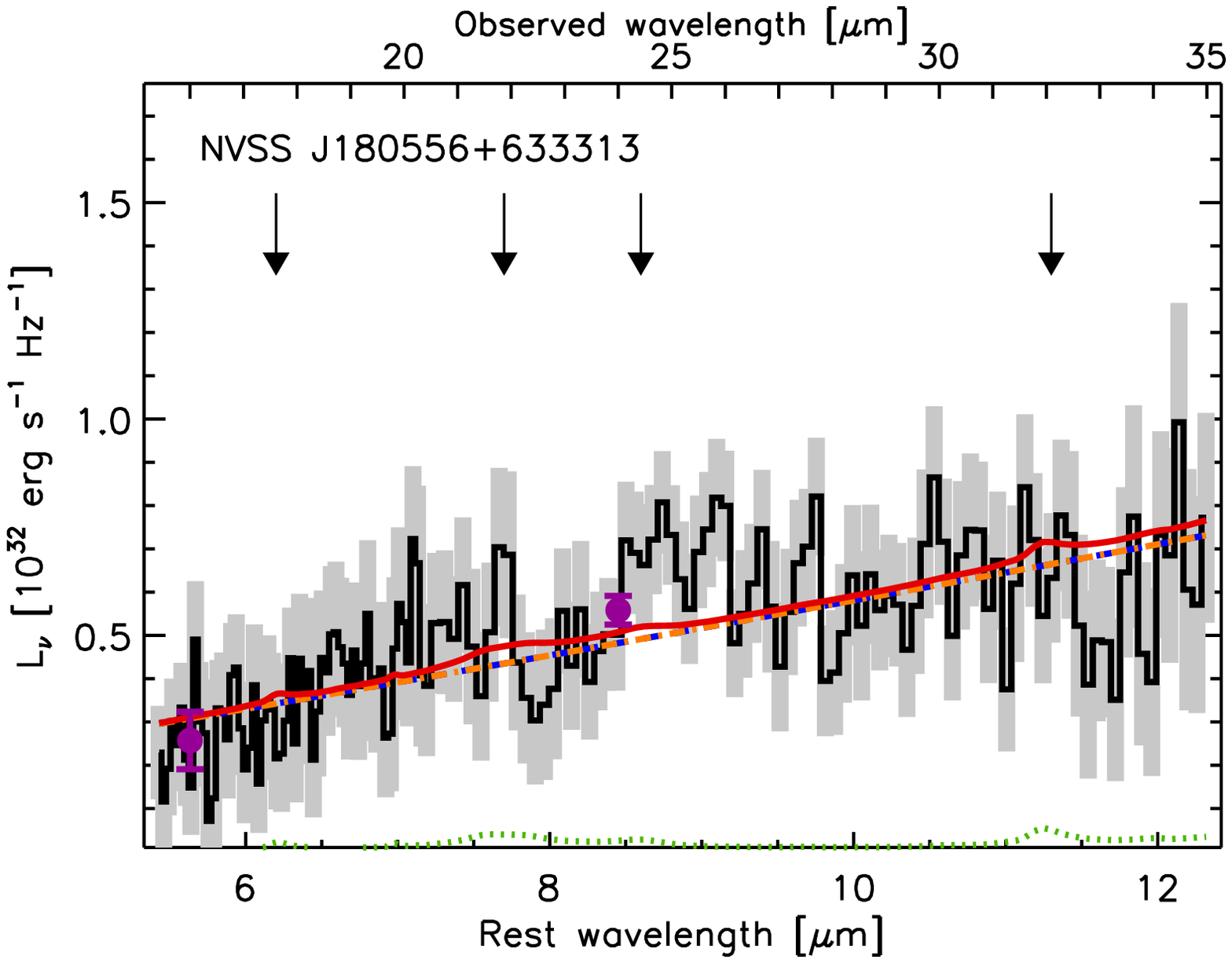,width=0.42\linewidth} &\epsfig{file=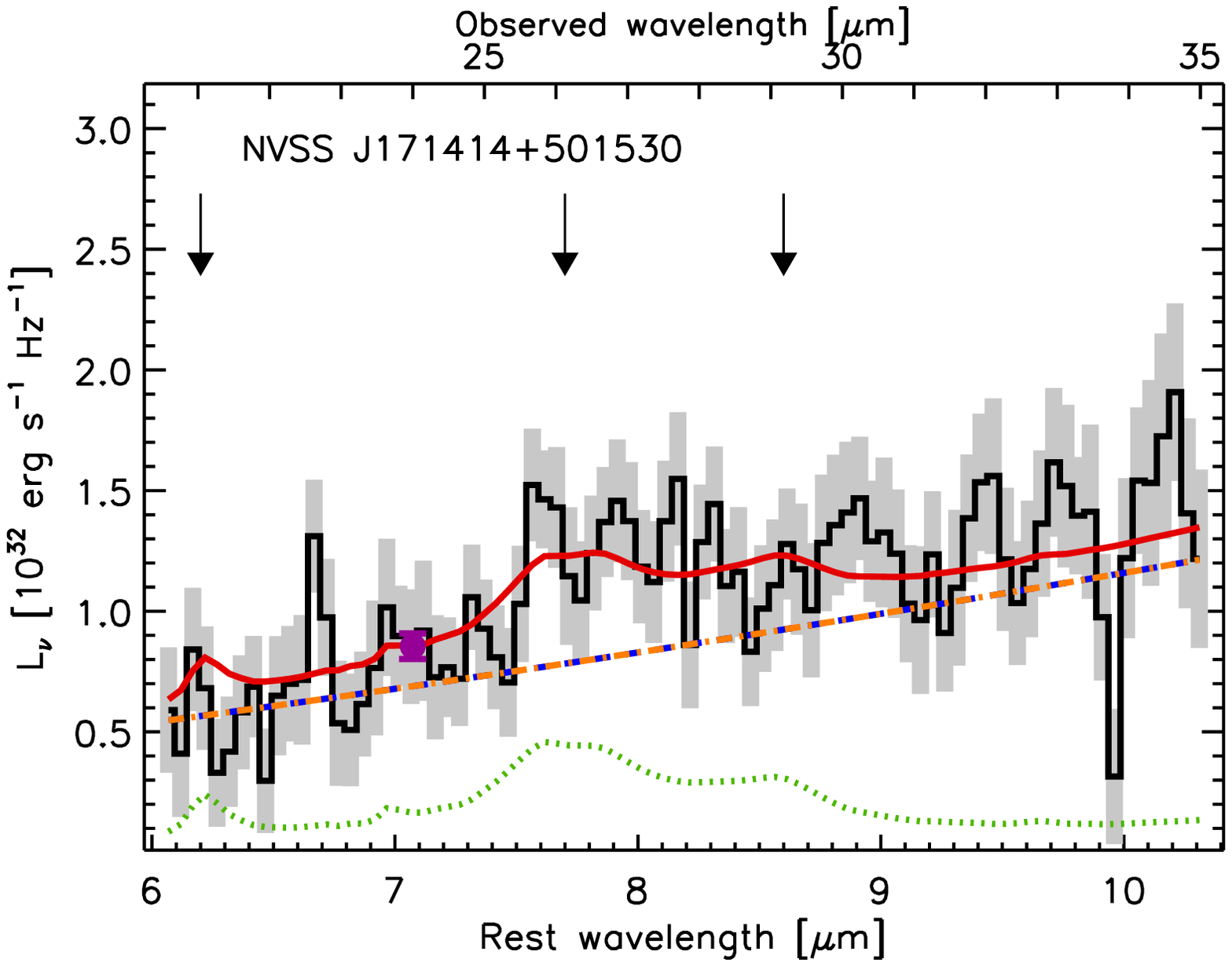,width=0.42\linewidth}\\
\epsfig{file=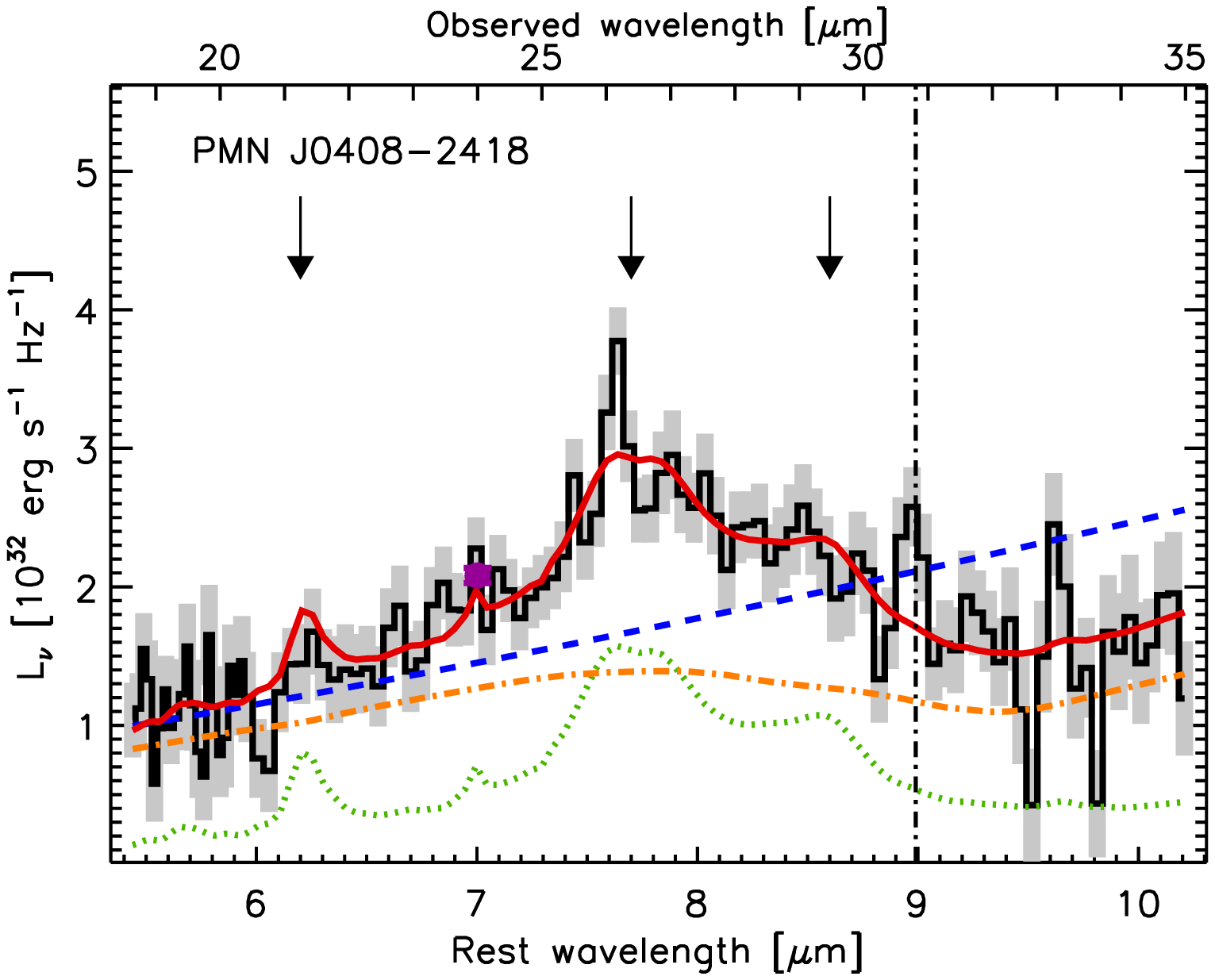,width=0.42\linewidth} &\epsfig{file=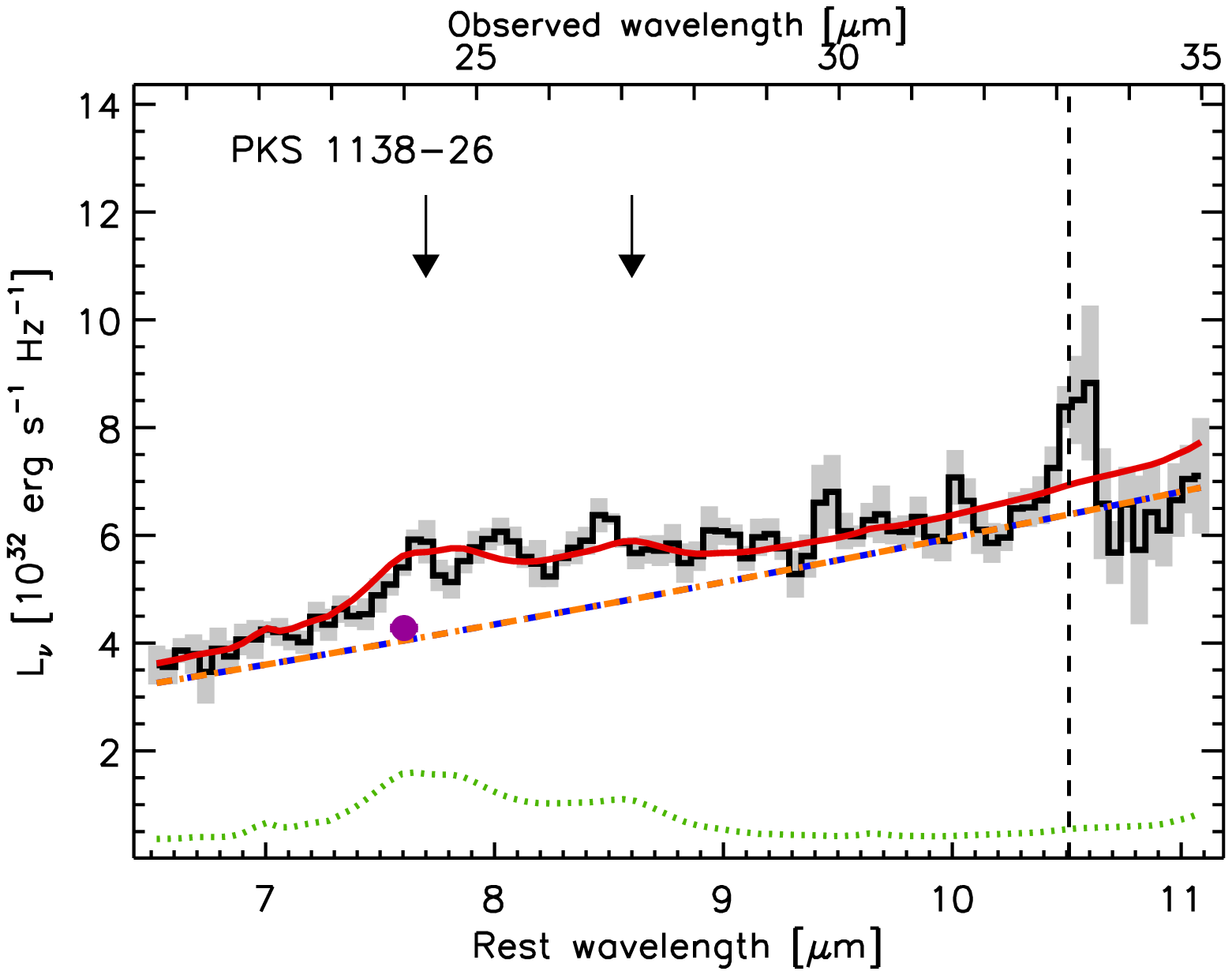,width=0.42\linewidth}\\
\epsfig{file=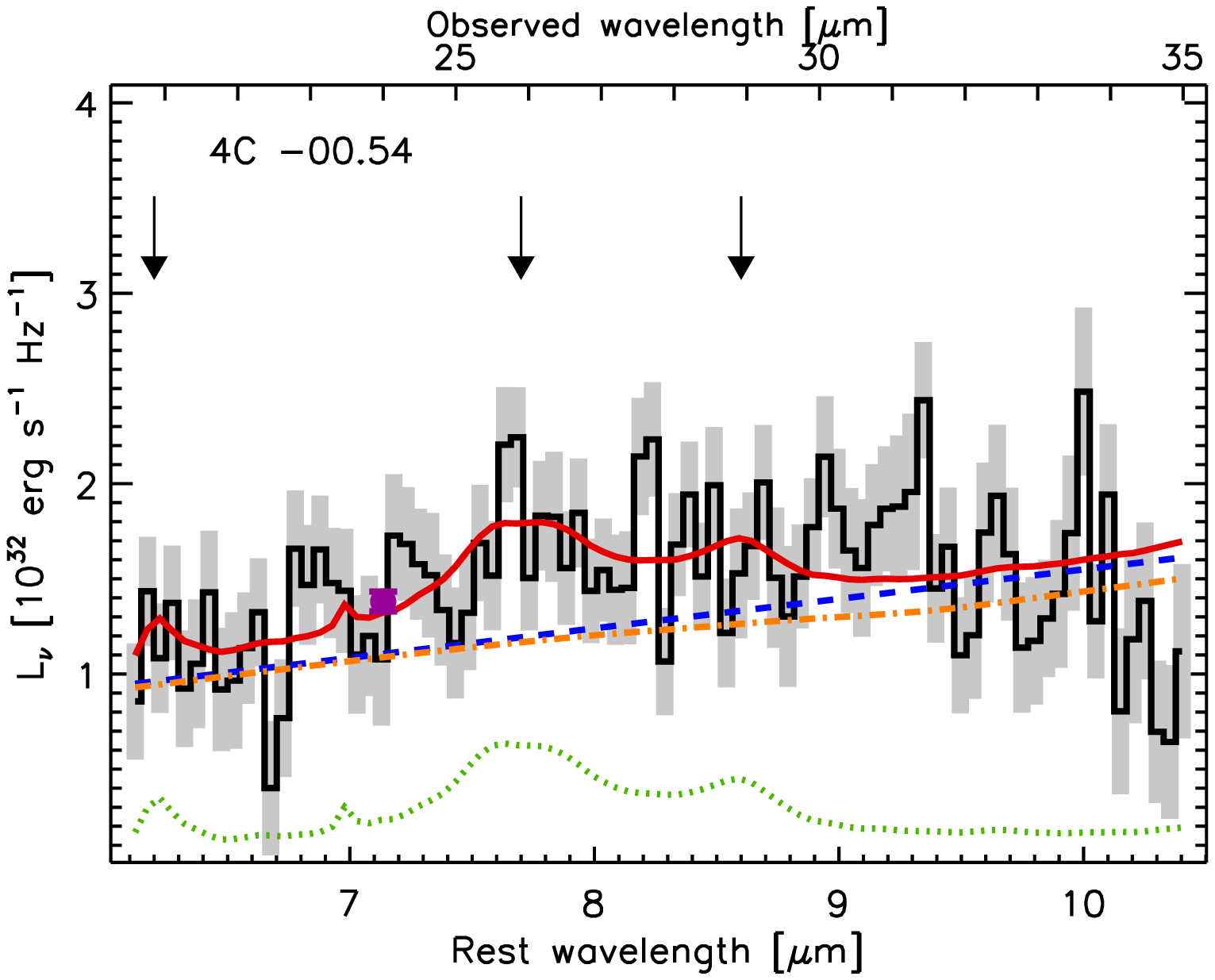,width=0.42\linewidth}\\
\end{tabular}
\caption{The mid-IR spectra of our sample along with their decomposition into AGN and host galaxy components. For all spectra, the data and the 1\,$\sigma$ uncertainties are represented by a histogram and shaded region, respectively, while the overall best-fitting is denoted by a solid red line, the host galaxy component is denoted by a dotted green line, while the power-law and extinguished power-law components are denoted by dashed blue and dot-dashed orange lines, respectively. Broadband MIPS/IRS photometry, though not included in the fitting, is also plotted (filled purple circles). The location of potential PAH features are indicated by arrows. The [S\,IV] emission line at 10.5\,$\mu$m is visible in the spectrum of 3C 239 and PKS 1138$-$26 while the [Ar\,III] line at 9\,$\mu$m is visible for PMN J0408$-$2418. The location of the [S\,IV] and [Ar\,III] lines in the three spectra are indicated by vertical dashed and dot-dashed lines, respectively.}
\label{Figure:spectra}
\end{figure*}

\begin{figure*}
\epsfig{file=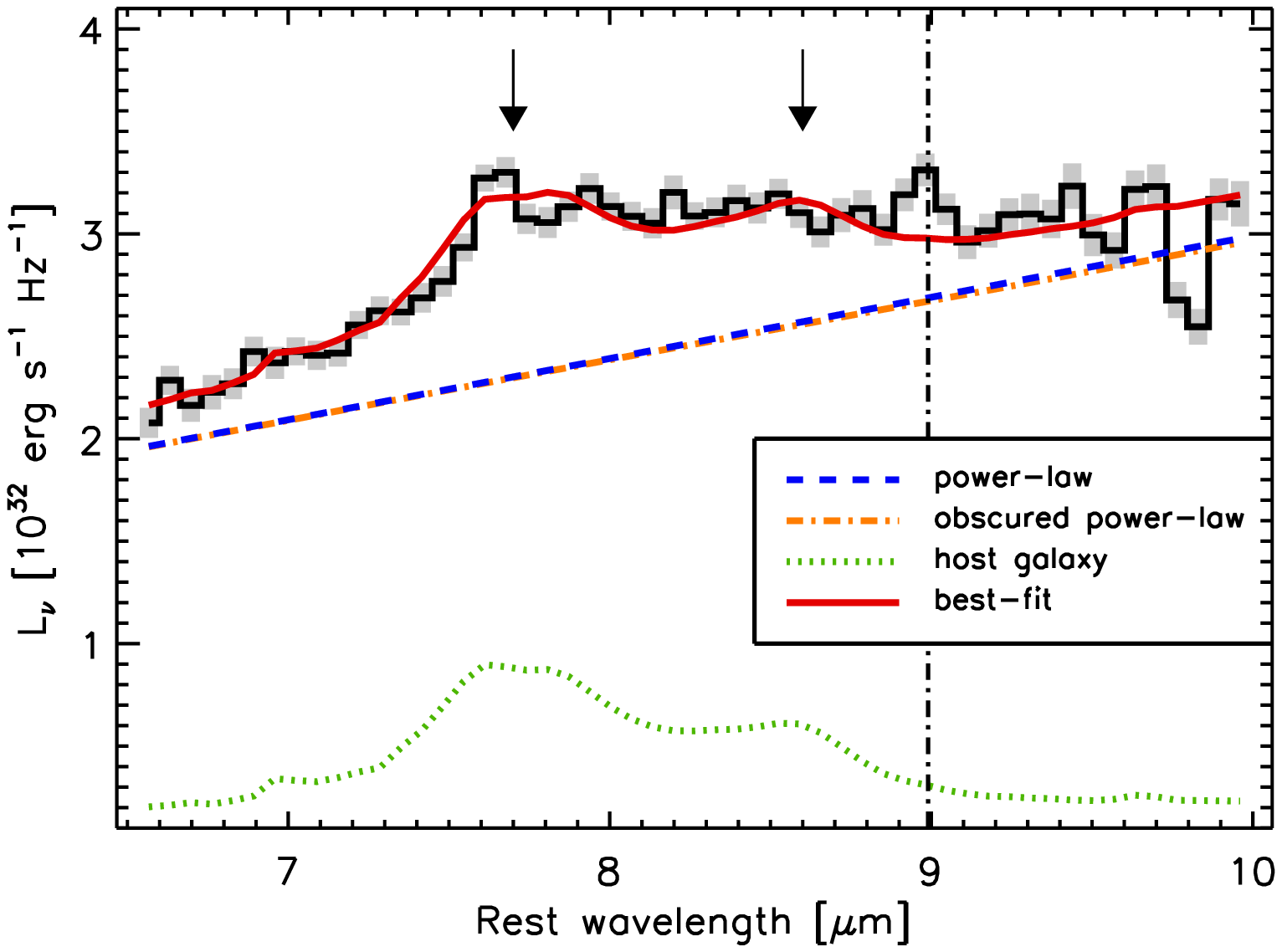,width=0.9\linewidth}
\caption{The mean spectrum of the seven radio galaxies in the sample, with the same fitted components as those in Figure \ref{Figure:spectra}. The location of potential PAH features are indicated by arrows. [Ar\,III] is detected at 9\,$\mu$m and its location is indicated by a dot-dashed line. The location of [S\,IV] lies outside the spectral coverage.}
\label{Figure:mean}
\end{figure*}

For each host galaxy template, $\chi^{2}$ minimisation is performed on the spectra, simultaneously solving for $k_{\rm 1}$, $k_{\rm 2}$, $\alpha$ and $N_{\rm H}$. In Figures \ref{Figure:spectra} and \ref{Figure:mean}, we show the resulting best-fitting components for the individual spectra and mean spectrum, respectively. To determine the confidence in the $\chi^{2}$ minimisation, the free parameters are allowed to deviate from their best-fitting values until the condition $\chi^{2}$\,-\,$\chi^{2}_{\rm min}$\,=\,$\Delta\chi^{2}$\,=\,a$^{2}$ is met; letting a\,=\,1 or 3 to find 1\,$\sigma$ and 3\,$\sigma$ uncertainties, respectively. The extent of the free parameter deviations from the best-fitting values represents their uncertainty limits. If $\Delta\chi^{2}$\,$<$\,9 is returned when the host galaxy component is not included, then at the 3\,$\sigma$ detection level, the fit is consistent with a scenario in which there is no star formation. 

\subsection{Deriving host galaxy and AGN properties}
\label{Section:deriving host galaxy and AGN properties section}
For the templates that fit best, we use the host galaxy normalisation factors to convert the 7.7\,$\mu$m PAH fluxes given in \citet{Brandl:06} into 7.7\,$\mu$m PAH luminosities, $L_{\rm 7.7 PAH}$, for our sample. The range of acceptable normalisation factors gives the range in PAH luminosity and thus, the uncertainty on the best-fit value. The PAH luminosities are then converted to total IR (8-1000\,$\mu$m) luminosities, $L_{\rm IR}$, (if assumed to be entirely due to star formation) through the conversion in P08 which is given as:
\begin{equation}
{\rm log}\,L_{\rm 7.7PAH}\,=\,(-1.5\,\pm\,0.1)\,+\,(0.95\,\pm\,0.01)\,{\rm log}\,L_{\rm IR}
\end{equation}
This conversion was determined empirically by P08, through the PAH and sub-mm luminosities of a combined sample of local starburst galaxies and high-redshift, star formation-dominated sub-mm galaxies (SMGs). These SMGs lie at similar redshifts to our souces.  $L_{\rm IR}$ is then converted to SFR via the \citet{Kennicutt:98} relation below, which assumes a starburst period $<$\,100\,Myr and a \citet{Salpeter:55} initial mass function (IMF):
\begin{equation}
\frac{\rm SFR}{\rm 1\,M_{\odot}\,yr^{-1}}\,=\,\frac{L_{\rm IR}}{\rm 5.8\,\times\,10^{9}\,L_{\odot}}
\end{equation}
Furthermore, the sample of P08 exhibit a 1\,$\sigma$ scatter of a factor of 2 about the derived $L_{\rm 7.7 PAH}-L_{\rm IR}$ conversion which translates to a range of SFR estimates from each derived PAH luminosity. The effect of this choice of conversion factor is investigated in \S\ref{Section:Spectral and fitting results} where we consider the conversion derived by MD09, in which a different IRS sample of SMGs was used.

From the AGN component, we determine monochromatic luminosities at rest-frame 6\,$\mu$m, $\nu L_{\rm \nu}$\,(6\,$\mu$m), where the continuum is less contaminated from spectral features than it is at longer wavelengths \citep{Lutz:04}. The spectrum for PKS 1138$-$26 and also the mean spectrum (due to reduced coverage) do not cover rest-frame 6\,$\mu$m and so when estimating their 6\,$\mu$m luminosities we assume the AGN emission at this wavelength is still characterised by the measured power-law and this component is simply extrapolated. In both cases, the extrapolation is $<$\,0.5\,$\mu$m. We also quantify the strength of silicate absorption at 9.7\,$\mu$m i.e. the 9.7\,$\mu$m apparent optical depth, by taking the natural log of the ratio of continuum emission, $f_{\rm cont}$, to observed emission, $f_{\rm obs}$, i.e. $\tau_{\rm 9.7\,\mu m}\,=\,ln(f_{\rm cont}/f_{\rm obs})\,\equiv\,N_{\rm H}\,\sigma_{\rm 9.7\,\mu m}$; $\sigma_{\rm 9.7\,\mu m}\,\approx$\,3.9\,$\times$\,10$^{-23}$\,cm$^{2}$. Note that we assume a clumpy rather than homogeneous distribution for the obscuring medium (see \S\ref{Section:discussion}) so our definition of optical depth is not equivalent to the traditional one \citep[e.g.][]{Nenkova:08a}.

\section{Results}
\label{Section:results}
\subsection{Spectral and fitting results}
\label{Section:Spectral and fitting results}
The individual spectra are presented in Figure \ref{Figure:spectra}. Note that the continua are consistent with the 16\,$\mu$m IRS blue peak-up and 24\,$\mu$m MIPS photometry. The mean S/N per resolution element for the whole sample ranges from $\sim$\,3-23. The mean spectrum has a mean S/N per resolution element of 37 and is shown in Figure \ref{Figure:mean}.

We find that all sources are AGN-dominated in the mid-IR, which is in agreement with the previous IR SED modelling of the whole SHzRG sample \citep[e.g.][DB10]{Seymour:07}. However, 6/7 also show a significant host galaxy contribution. For the other source (NVSS J180556$+$633313), a fit that includes no host galaxy component differs from the best-fitting model by $\Delta$$\chi^{2}$\,$<$\,1, so we do not significantly detect star formation in this source. For the radio galaxies whose spectra have a mean S/N per resolution element $<\,10$ (NVSS J180556$+$633313, NVSS J171414$+$501530 and 4C -00.54), more than one host galaxy template can be fitted with $\Delta\chi^{2}\,<$\,1. In this case, similar 7.7\,$\mu$m PAH luminosities are derived from the acceptable templates and so although the PAH luminosities have been derived from a model-fitting process, they are at least insensitive to the choice of host galaxy template.

At the extreme, we calculate very high SFRs of $\sim$\,1000\,M$_{\odot}$\,yr$^{-1}$ (for 4C 23.56, PMN J0408$-$2418 and PKS 1138$-$26). We find an SFR of 620\,$^{+\,620}_{-\,310}$\,M$_{\odot}$\,yr$^{-1}$ for the mean spectrum and a mean SFR of 610\,$\pm$\,180\,M$_{\rm \odot}$\,yr$^{-1}$ for the individual sources. The error of the mean SFR is derived from the scatter of the individual SFRs and does not take into account the PAH to SFR conversion uncertainty. We examine how the radio galaxy with the highest equivalent width 7.7\,$\mu$m PAH feature in its spectrum, PMN J0408$-$2418, contributes to the mean SFR by creating a mean spectrum without including the source. We derive a mean SFR of 510\,$^{+\,520}_{-\,260}$\,M$_{\odot}$\,yr$^{-1}$ suggesting that star formation is abundant in our sample. Except for NVSS J180556$+$633313, the total IR luminosities (inferred from 7.7\,$\mu$m PAH luminosities) alone qualify these objects as ultra luminous infra-red galaxies (ULIRGs; 12\,$\leq$\,log(L$_{\rm IR}$/L$_{\rm \odot}$)\,$<$\,13). Assuming the IR-radio correlation for star-forming galaxies holds at high-redshift \citep{Seymour:09, Huynh:10}, we calculate star-forming radio luminosities from the derived SFRs using the \citet{Bell:03} conversion factor. The star-forming radio luminosities are $\sim$\,5 orders of magnitude lower than the measured radio luminosities and so the contribution of star formation to the radio emission in these HzRGs is negligible as expected. We next investigate the effect our choice of the P08 $L_{\rm 7.7 PAH}-L_{\rm IR}$ conversion has on our results. Applying the MD09 conversion (see \S\ref{Section:deriving host galaxy and AGN properties section}) to the mean spectrum we find a very similar SFR of 710\,$\pm$\,40\,M$_{\rm \odot}$\,yr$^{-1}$, well within the 1\,$\sigma$ uncertainties using the P08 conversion. No 1\,$\sigma$ scatter in the relation is given by MD09 and so the uncertainty given here reflects only the uncertainty in the PAH luminosity. The similarity between the SFRs from the two conversion factors is an indication that our SFRs are not sensitive to the choice of conversion factor from PAH luminosity. To derive their $L_{\rm 7.7 PAH}-L_{\rm IR}$ relations, both P08 and MD09 used similar methods and objects to us, in that their mid-IR spectra were decomposed into AGN and host galaxy components. Therefore, it is appropriate to apply these conversions for our sources.

Using the stellar masses from Table \ref{Table:source_info} and the SFRs from Table \ref{Table:fitting_results}, we now calculate the specific star formation rate (sSFR) of our sample. The sSFR is a measure of a galaxy's rate of growth with respect to its stellar mass. We do this for the two sources that have stellar mass estimates and we find lower limits for the other sources that only have upper limits for their stellar mass (due to unknown contributions from the AGN to the \textit{H}-band luminosities, from which the masses are derived). The exception to this is NVSS J180556$+$633313, for which we only have upper limits for the SFR and stellar mass and therefore are unable to calculate its sSFR. It is noteworthy that different IMFs have been used to calcuate the stellar masses and SFRs (see \S\ref{Section:deriving host galaxy and AGN properties section}). The \citet{Kroupa:01} IMF underestimates the total stellar mass compared to the \citet{Salpeter:55} IMF, for a galaxy of a given luminosity, by a factor of $\sim$\,1.5 (Karczewski et al. 2012; \textit{in prep.}). We account for this factor when calculating the sSFRs for our sample. We derive sSFRs of 0.31\,$^{+\,0.32}_{-\,0.17}$\,Gyr$^{-1}$ and 3.12\,$^{+\,3.14}_{-\,1.58}$\,Gyr$^{-1}$ for 3C 239 and PMN J0408$-$2418, respectively, while three of the other four galaxies have sSFRs $>\,1$\,Gyr$^{-1}$. 

From the AGN component, we find very high mid-IR luminosities, with $\nu L_{\rm \nu}$\,(6\,$\mu$m)\,$\approx$\,10$^{46}$\,erg\,s$^{-1}$ for the mean spectrum and $\nu L_{\rm \nu}$\,(6\,$\mu$m) between 10$^{45.2}$ and 10$^{46.5}$\,erg\,s$^{-1}$ for the individual objects. We also find that similar AGN luminosities were derived when each template was fitted, showing that the AGN component is not sensitive to the choice of template. As for the obscuration of the AGN component, we find low to moderate 9.7\,$\mu$m optical depths ($\tau_{\rm 9.7\,\mu m}\,<$\,0.8) for the sample. When investigating the degeneracy of $\alpha$ and extinction, we find little covariance of the 1\,$\sigma$ confidence regions in the parameter space and so we observe little degeneracy.

Although we find many of these radio galaxies have significant star formation, one important point to note is that the sample could be biased towards objects with significant star formation because, at $z\,\sim$\,2, the 7.7\,$\mu$m PAH feature shifts into the 24\,$\mu$m MIPS band. While these radio galaxies were not 24\,$\mu$m selected, they are all bright at 24\,$\mu$m (S$_{\rm 24\mu m}$\,$>$\,0.4\,mJy) as this was a requirement for IRS observations. This means a substantial fraction of the 24\,$\mu$m emission could be from star formation. In Table \ref{Table:fitting_results}, we consider the contribution of the host galaxy component to the total flux in the 24\,$\mu$m MIPS passband for our sample. For approximately half the sample, these contributions are significant ($\gtrsim$\,20\,\%) but only one source would potentially drop out of the sample if the star-forming contributions were subtracted from the 24\,$\mu$m flux densities. Therefore, these sources are mainly selected on their AGN power but the star-forming bias may still exist.

\subsection{Average star formation rate of the HzRG population}
As mentioned previously (in \S\ref{Section:sample selection}), the original SHzRG sample is representative of the HzRG population. In addition, amongst the sources that are within our redshift range and bright at 24\,$\mu$m (S$_{\rm 24\mu m}$\,$>$\,0.4\,mJy) i.e. the sources within the region enclosed by the red dashed lines in Figure \ref{Figure:sample}, we can treat those selected for IRS observations as a random selection. As a result, even though we have estimated the SFR of only seven HzRGs, we can estimate the total mean SFR for the overall HzRG population in our redshift range ($1.5\,<\,z\,<\,2.6$). This is possible with a few assumptions. We can assume the mean SFR ($\langle$\,SFR\,$\rangle$\,=\,610\,$\pm$\,180\,M$_{\rm \odot}$\,yr$^{-1}$) of our sample is equal to the mean SFR of the sources within the region enclosed by the red dashed lines in Figure \ref{Figure:sample}. We can also make two assumptions on the SFR of the sources in our redshift that are `faint' at 24\,$\mu$m (S$_{\rm 24\mu m}$\,$<$\,0.4\,mJy) i.e. the sources within the region enclosed by the black dashed lines in Figure \ref{Figure:sample}. As in \citet{Seymour:11}, firstly, to obtain a lower limit to the total mean SFR, we assume that the mid-IR faint sources have SFRs of zero. The total mean SFR then just scales with the mean SFR of the mid-IR-bright sources by the fraction of mid-IR bright sources (i.e. total mean SFR\,=\,24/30\,$\times$\,$\langle$\,SFR\,$\rangle$\,=\,490\,M$_{\rm \odot}$\,yr$^{-1}$). Secondly, to obtain an upper limit, we assume that the mid-IR faint sources have SFRs equal to the mean SFR of the mid-IR bright sources. With this approach, we constrain the total mean SFR of the HzRG population in our redshift range to be 490-610\,M$_{\rm \odot}$\,yr$^{-1}$. Taking into account the standard error of the mean of $\pm$\,30\% gives the range 340-790\,M$_{\rm \odot}$\,yr$^{-1}$.

\subsection{Relating AGN to host galaxy parameters}
We now investigate the relationship between AGN and host galaxies, conducting Spearman's rank statistics to identify possible correlations for the physical parameters of our sample. More specifically, we determine the Spearman's rank coefficient, $\rho$, and the significance of the deviation of $\rho$ from zero, $\sigma$. Values very close to unity for $\rho$ and $\sigma$ would indicate a significant correlation exists.

\begin{figure}
\epsfig{file=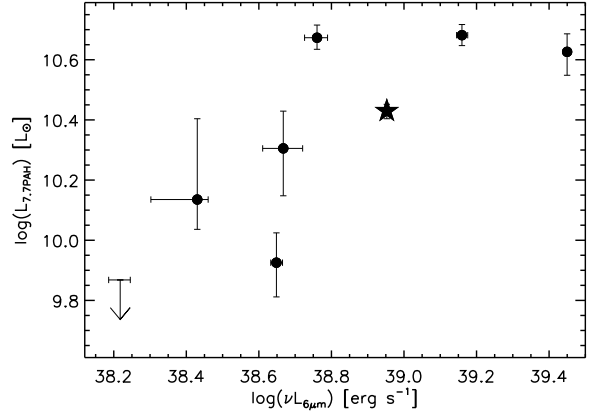,width=1\linewidth}
\caption{7.7\,$\mu$m PAH luminosity as a function of mid-IR AGN luminosity for the sample. The mean spectrum is represented by a star.}
\label{Figure:pahlum_agn}
\end{figure}
 
\begin{figure*}
\epsfig{file=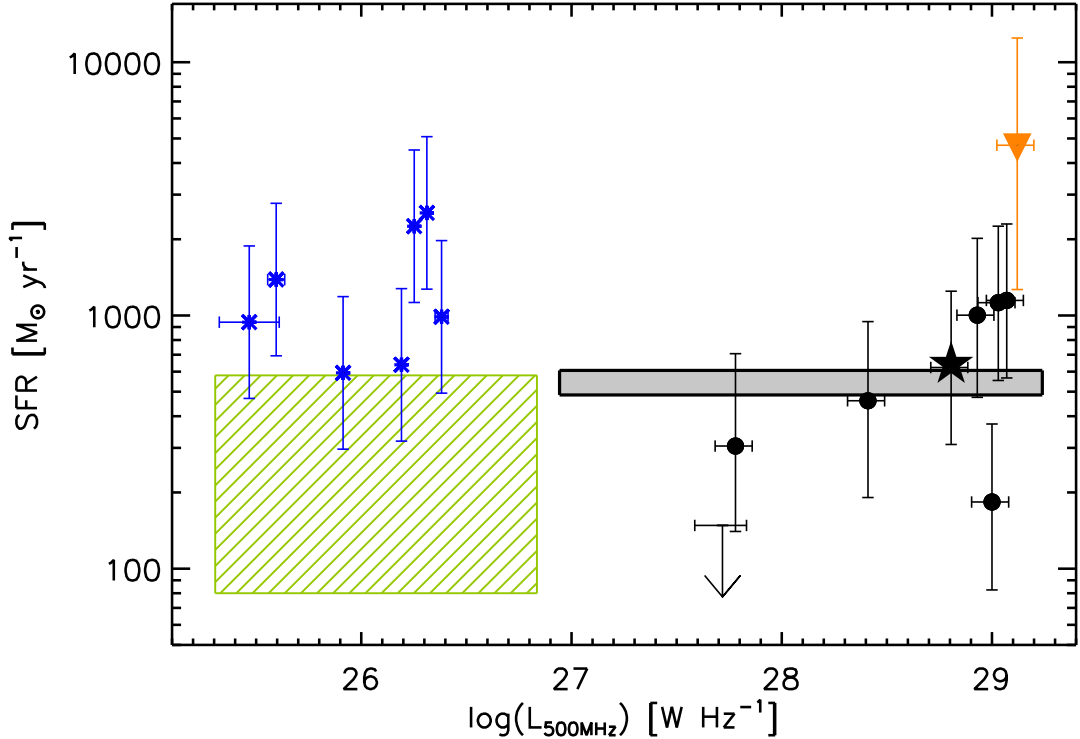,width=0.86\linewidth}	
\caption{SFR as a function of 500\,MHz luminosity for our sources (filled black circles), the high-redshift radio-loud AGN from \protect \citet[blue asterisks]{Sajina:07b} and TXJ 1908+7220 \citep[$z$\,=\,3.53, orange triangle,][]{Seymour:08}. Also plotted is the mean SFR for the 30 SHzRGs objects in our redshift range (grey region) and the parameter space covered by the \protect \citet{Seymour:11} high-redshift (1.2\,$<\,z\,<$\,3) sample of radio-loud AGN (green hashed region). For the \protect \citet{Sajina:07b} sources, 500\,MHz luminosities are calculated from 1.4\,GHz luminosities using their radio spectral indicies. For TXJ 1908+7220, we use the 500\,MHz luminosity from DB10. For the \protect \citet{Seymour:11} sources, 500\,MHz luminosities are calculated from 1.4\,GHz luminosities by assuming a radio spectral index of -0.75. For the \protect \citet{Sajina:07b} sources and TXJ 1908+7220, SFRs are calculated from 7.7\,$\mu$m PAH luminosities using the same method as our sample. The mean spectrum is represented by a star.}
\label{Figure:sfr_radio}
\end{figure*} 

In Figure \ref{Figure:pahlum_agn} we plot 7.7\,$\mu$m PAH luminosity against rest-frame 6\,$\mu$m AGN luminosity for all our sources and the mean spectrum. The scales for both parameters vary by more than an order of magnitude, showing a large diversity in the limited sample and we find only a marginal correlation ($\rho$\,=\,0.67, $\sigma$\,=\,0.90). Next we plot SFR versus 500\,MHz luminosity for our sample and the mean spectrum in Figure \ref{Figure:sfr_radio} along with results from the literature. We also show the mean SFR for the 30 SHzRGs objects in our redshift range (grey region), the high-redshift radio-loud AGN from \citet[blue asterisks]{Sajina:07b}, the higher-redshift radio galaxy TXJ 1908+7220 \citep[$z$\,=\,3.53, orange triangle,][]{Seymour:08} and the range in mean SFR and 500\,MHz luminosity for the \citet{Seymour:11} high-redshift (1.2\,$<\,z\,<$\,3) sample of radio-loud AGN (green hashed region). It is visible that mid-IR bright radio-loud AGN over a few orders of magnitude in radio luminosity can have significant star formation.

\begin{figure}
\epsfig{file=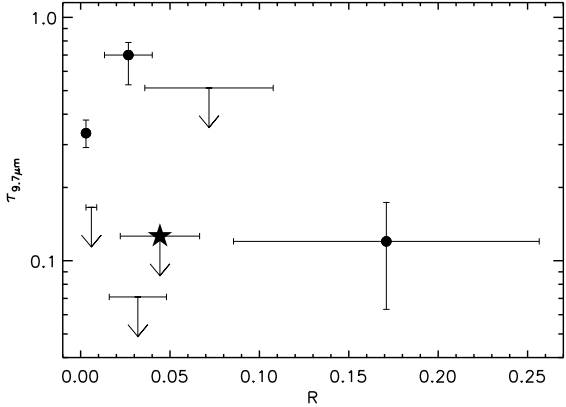,width=1\linewidth}
\caption{9.7\,$\mu$m optical depth as a function of radio core dominance. The mean spectrum is represented by a star.}
\label{Figure:ext}
\end{figure}

Radio core dominance, R, defined as the ratio of core and extended radio flux at rest-frame 20\,GHz, is a measure of the extent to which the central component of the radio source contributes to the total radio emission. It can therefore be used as a proxy for the orientation of the radio jets \citep[Drouart et al. 2012; \textit{in prep.}]{Kapahi:82}. Using the radio core dominances from Table \ref{Table:source_info}, we next consider if jet orientation shows any relationship with the mid-IR obscuration of these AGN. Due to anisotropic, relativistic beaming effects, the cores of the radio sources will appear brighter, relative to their jets, if the jets are orientated close to the line of sight. In Figure \ref{Figure:ext} we plot 9.7\,$\mu$m optical depth versus the radio core dominance for our sample and observe no correlation. A correlation may have been anticipated as AGN mid-IR obscuration and jet orientation are expected to correlate if their axes are perpendicularly alligned. The likely reason for observing no correlation is discussed in \S\ref{Section:discussion}.

\subsection{Other spectral features}
The rest-frame 10.5\,$\mu$m [S\,IV] atomic emission line is detected in the spectra of 3C 239 and PKS 1138$-$26, with equivalent widths (EWs) of 0.07\,$\pm$\,0.01\,$\mu$m and 0.04\,$\pm$\,0.01\,$\mu$m, respectively. For PMN J0408$-$2418, the rest-frame 9\,$\mu$m [Ar\,III] emission line is detected with an EW of 0.06\,$\pm$\,0.01\,$\mu$m. We also detect [Ar\,III] in the mean spectrum (EW\,=\,0.015\,$\pm$\,0.003\,$\mu$m) but the location of [S\,IV] lies outside the spectral coverage. These EWs are larger than those found for a sample of local Seyfert galaxies \citep[EW$_{\rm [S\,IV]}$\,$\lesssim$\,0.03\,$\mu$m, EW$_{\rm [Ar\,III]}$\,$\lesssim$\,0.01\,$\mu$m,][]{Honig:08} and are comparable to those for other powerful type-2 AGN \citep{Shan:12}. No H$_{\rm 2}$ emission is detected in any of our spectra. However, for PKS 1138$-$26, H$_{\rm 2}$ is detected in the higher S/N IRS spectrum presented in \cite{Ogle:12}. Therefore, the non-detections in our spectra may be partly due to low S/N.

\section{Discussion}
\label{Section:discussion}
Through the detected PAH emission in their mid-IR spectra, we observe that some HzRGs are forming stars at a very high rate. We find a high mean SFR of 340-790\,M$_{\rm \odot}$\,yr$^{-1}$ for the overall HzRG population in the redshift range $1.5\,<\,z\,<\,2.6$. However, not all HzRGs at this epoch have such significant star formation, as is the case for NVSS J180556$+$633313 (SFR\,=\,20\,$^{+\,50}_{-\,20}$\,M$_{\rm \odot}$\,yr$^{-1}$). High SFRs have also been detected in other high redshift radio-loud AGN, for example, the AGN-dominated ULIRGs of \citet[SFR\,$\approx$\,1000\,M$_{\rm \odot}$\,yr$^{-1}$]{Sajina:07b}. However, these sources were selected to be bright in the mid-IR (S$_{\rm 24\mu m}$\,$\gtrsim$\,0.9\,mJy) before they were discovered to be radio-loud, implying that the sample might have been biased towards objects with significant star formation. As shown in Figures \ref{Figure:pahlum_agn} (right panel) and \ref{Figure:sfr_radio}, the \citet{Sajina:07b} sample have lower radio luminosities than our sources. Therefore, PAH emission exists in mid-IR bright radio-loud AGN over $\sim$\,4 orders of magnitude in radio luminosity. In Figure \ref{Figure:sfr_radio} we also plot the less-luminous high-redshift AGN from \citet{Seymour:11}, which have a mean SFR of 80-580\,M$_{\rm \odot}$\,yr$^{-1}$, comparable to or less than the mean SFR for the HzRG population in our redshift range. For our sample, we have not found a significant correlation between star formation and AGN power, for both mid-IR and radio luminosities, of powerful radio-loud AGN at high redshift. When including the \citet{Sajina:07b} and \citet{Seymour:11} samples, the scatter increases. This is consistent with \citet{Mullaney:11} who also examined SFR as a function of AGN power. However, other studies have been able to show a connection between star formation and central black hole accretion \citep{Shao:10, Bonfield:11}.

\begin{figure}
\epsfig{file=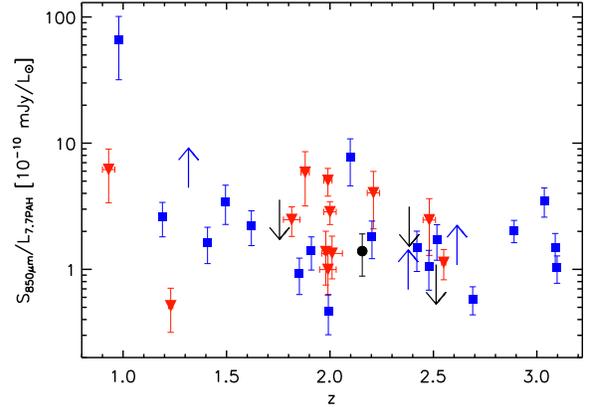,width=1\linewidth}
\caption{S$_{\rm 850\,\mu m}$/$L_{\rm 7.7 PAH}$ verses redshift for the sources in our sample with sub-mm observations (black circle/upper limits) and the sub-mm galaxies from P08 (red triangles) and MD09 (blue squares/lower limits). The lower limits for some of the MD09 sources are due to greater contributions from noise in the 7.7\,$\mu$m PAH region.}
\label{Figure:flux_rat_z}
\end{figure}

To further investigate whether the use of the $L_{\rm 7.7 PAH}-L_{\rm IR}$ conversion from P08 was appropriate for our sources, we can compare the PAH and sub-mm emission from our sources to the emission from the IRS-observed SMGs that were used to (partly) derive the conversion. As mentioned previously, four radio galaxies in our sample have sub-mm observations; three have non-detections \citep[3C 239, S$_{\rm 850\,\mu m}$\,=\,0.00\,$\pm$\,1.00\,mJy; 4C +23.56, S$_{\rm 850\,\mu m}$\,=\,1.68\,$\pm$\,0.98\,mJy and NVSS J171414$+$501530, S$_{\rm 850\,\mu m}$\,=\,0.99\,$\pm$\,1.10\,mJy; corrected for radio synchrotron emission contamination;][]{Archibald:01} and the other is detected \citep[PKS 1138$-$26, S$_{\rm 850\,\mu m}$\,=\,6.7\,$\pm$\,2.4\,mJy;][]{Stevens:03}. For these sources and the SMGs from P08 and MD09 that had full spectral coverage of the 7.7\,$\mu$m PAH feature (12/13 and 21/24 objects, respectively), we plot the ratio of 850\,$\mu$m flux density to 7.7\,$\mu$m PAH luminosity as a function of redshift in Figure \ref{Figure:flux_rat_z}. Due to the strong negative \textit{K}-correction at 850 $\mu$m for thermal dust emission, the 850\,$\mu$m flux density can be used as a crude proxy for far-IR luminosity at $z\,>\,1$. We retrieved 850\,$\mu$m flux densities for the SMGs from \citet{Pope:06}, \citet{Chapman:01} and \citet{Chapman:05}, while redshifts were either derived from the IRS data by P08 and MD09 or optical spectroscopic redshifts were retrieved from \citet{Chapman:05}. We find our radio galaxies have S$_{\rm 850\,\mu m}-L_{\rm 7.7 PAH}$ ratios (upper limits for the sub-mm non-detections) within the ranges of both samples of SMGs and so this indicates that it was reasonable to apply the $L_{\rm 7.7 PAH}-L_{\rm IR}$ P08 conversion to our sources.

We turn now to the 9.7\,$\mu$m silicate features in the IRS spectra. We begin by examining whether the host galaxy could contribute significantly to the 9.7\,$\mu$m silicate feature by considering the emission at shorter wavelengths where the effects of extinction are greater than in the mid-IR. Near-IR (rest-frame optical) spectra exist for five sources in our sample; namely 4C 23.56 and 4C$-$00.54 \citep{Humphrey:08}, PKS 1138$-$26 \citep{Nesvadba:06, Humphrey:08}, NVSS J171414$+$501530 \citep{Motohara:01} and PMN J0408$-$2418 \citep{McCarthy:92}. Only one source (PKS 1138$-$26) has a significant broad component to H$\alpha$, from which \cite{Nesvadba:06} estimated the total extinction in the V-band, A$_{\rm V}$. They find A$_{\rm V}$\,=\,8.7, and this source has a very weak silicate feature ($\tau_{\rm 9.7\,\mu m}$\,$<$\,0.07 at 3\,$\sigma$). Whether the host galaxy contributes to the feature or not, any silicate absorption from the torus must be weak ($\tau_{\rm 9.7\,\mu m}$\,$<$\,0.8) because absorption from the host galaxy can only increase the depths of the features. From the narrow-line H$\alpha$\,/\,H$\beta$ ratios, \citet{Humphrey:08} placed constraints on the A$_{\rm V}$ of the extended emission line region (EELR) for 4C 23.56, PKS 1138$-$26 and 4C$-$00.54. These regions are external to the torus \citep{Fosbury:82, Fosbury:84, Baum:88, Tadhunter:89} so any obscuration is likely to be due to the host galaxy. Both 4C 23.56 and PKS 1138$-$26 have narrow-line H$\alpha$\,/\,H$\beta$ ratios consistent with case B photoionisation with no extinction. A greater H$\alpha$\,/\,H$\beta$ ratio was derived for 4C$-$00.54, suggesting non-negligible extinction from the host galaxy. From the ratios, \citet{Humphrey:08} derived A$_{\rm V}$\,$<$\,1.2 and $<$\,0.7 for 4C 23.56 and PKS 1138$-$26 respectively, and A$_{\rm V}$\,$>$\,1.2 for 4C$-$00.54. We translate A$_{\rm V}$ to silicate absorption strength of the EELR, $\tau_{\rm 9.7\,\mu m}$(EELR), using the same extinction curve used in the fitting and find $\tau_{\rm 9.7\,\mu m}$(EELR)\,$<$\,0.09, \,$<$\,0.05 and \,$>$\,0.07 for 4C 23.56, PKS 1138$-$26 and 4C$-$00.54 respectively. These values are consistent with the absorption strengths from the fitting of $\tau_{\rm 9.7\,\mu m}$\,=\,0.12\,$\pm$\,0.06, $<$\,0.07 and $<$\,0.51 (3\,$\sigma$ upper limits). Regarding the near-IR spectra for NVSS J171414$+$501530 and PMN J0408$-$2418, narrow H$\alpha$ emission is detected but there is no coverage of the H$\beta$ line and so we cannot constrain the host galaxy extinction for these sources. Considering the sample overall, we cannot rule out a contribution from the host galaxy extinction to the silicate absorption features, but neither can we identify any silicate absorption features for which we are confident that the host galaxy contributes.

To characterise the tori in our sources, we next consider the attempts that have been made to model the torus geometry using the 9.7\,$\mu$m silicate feature. It has been shown that both homogeneous \citep{Pier:92} and clumpy \citep{Levenson:07, Nenkova:08b} torus models can synthesise the silicate feature observed for a number of type-2 AGN (\citealt{Hao:07, Spoon:07}; \citealt{Nikutta:09}). However, a homogeneous geometry is unphysical for the torus due to thermal spallation \citep{Krolik:88} and although such a geometry is a reasonable approximation as long as the distance between the clouds is small in comparison to the scale of the whole structure \citep{Pier:92}, we consider only a clumpy distribution from here on in. \citet{Nenkova:08a} showed that for a clumpy torus, a silicate feature in emission could be produced, particularly if there are very few clouds along any radial equatorial ray. Such emission from the torus could potentially be masked by stronger silicate in absorption from the host galaxy. We cannot rule out such a situation from the data but as we have no evidence to support this, we make the caveat that in the following discussion we attribute the silicate features (or lack of) entirely to the torus component, in line with other similar studies \citep[e.g.][]{Levenson:07, Spoon:07, Sajina:07a, Cleary:07}. Considering the model from \citet{Nenkova:08a}, we constrain some of the physical parameters of the torus for our AGN. For instance, \citet{Nenkova:08a} showed that a soft-edged distribution better matches observations of the nuclear emission of a sample of Seyfert galaxies than a sharp-edged geometry does (see their Figure 1 for a schematic diagram of the two toroidal distributions). Furthermore, their models were able to reproduce the 9.7\,$\mu$m silicate feature properties of Seyfert galaxies when the average number of clouds along any radial equatorial ray is $\sim$\,5-15, the angular width of the torus is $\sim$\,15-45\,$^{\circ}$ and the actual optical depth of the individual clouds is $\sim$\,30-100. Considering PMN J0408$-$2418; the one HzRG for which we calculate a moderate 9.7\,$\mu$m optical depth ($\tau_{\rm 9.7\,\mu m}$\,=\,0.70\,$^{+\,0.09}_{-\,0.17}$), for the torus model to reproduce this absorption strength, an average of $\sim$\,10-15 clouds along any radial equatorial ray are required, with each cloud having an optical depth of $\sim$\,10-80. An angular width of $\sim$\,30-45\,$^{\circ}$ is also needed. This particular model has a radial distribution profile of r$^{-2}$ and an outer-to-inner radius ratio of 30. A model with an angular width as low as 15\,$^{\circ}$ can reproduce the absorption strength but for this a cloud optical depth of $\sim$\,20-50 is required. For the other HzRGs that have weaker 9.7\,$\mu$m optical depths, the silicate absorption strengths can be reproduced by a model that has an angular width of $\sim$\,15-45\,$^{\circ}$ and also a radial distribution profile of r$^{-2}$ and an outer-to-inner radius ratio of 30. However, we cannot differentiate a model that has a lower number of clouds, $\sim$\,5, and a moderate cloud optical depth, $\sim$\,40, with a model that has a higher number of clouds, $\sim$\,10-15, and either a much lower or much higher cloud optical depth, $\sim$\,10 or $\sim$\,100. This is because both the number of clouds and the cloud optical depth are the major driving forces behind the strength of silicate absorption \citep[see Figure 6 in][]{Nenkova:08a}, implying a higher level of degeneracy in these fitting parameters when modelling the torus of AGN with weaker optical depths.

To investigate the connection between AGN mid-IR obscuration and orientation, we compare our results to those of \citet{Cleary:07}. For an orientation-unbiased sample of lower-redshift ($z<1$) radio galaxies, they find an anti-correlation between 9.7\,$\mu$m optical depth and radio core dominance which implies there is a coupling of the axes of the radio jets and the mid-IR obscuration. Although we observe no correlation between these parameters in Figure \ref{Figure:ext}, for the same range in core dominance, our sources cover a similar range in optical depth. The reason we do not observe a correlation is likely to be because the smaller sample of our study means we probe a smaller region in parameter space which makes it more difficult to observe the correlation. Comparing to other studies, our 9.7\,$\mu$m optical depths are similar to those for radio-quiet type-2 AGN \citep{Hao:07, Spoon:07} and other lower-redshift radio galaxies \citep{Haas:05, Ogle:06}. This suggests that the tori in these AGN have similar properties to the tori in our sample. However, on average, higher levels of obscuration ($\tau_{\rm 9.7\,\mu m}$\,$>$\,1) have been found for other samples, such as the radio-loud AGN/ULIRGs of \citet{Sajina:07b} and local Compton-thick ($N_{\rm H}\,>\,1.5\,\times\,10^{24}$\,cm$^{-2}$) AGN \citep{Goulding:12}. For \citet{Sajina:07b}, this is likely due to their sources being selected to have large mid-IR-optical colours. The authors from both studies attribute their larger optical depths to host-galaxy contribution to the obscuration of the AGN by way of acting as a cold dust screen.

We next examine these radio galaxies in the context of galaxy formation and evolution. The sSFR of 0.31\,$^{+\,0.32}_{-\,0.17}$\,Gyr$^{-1}$ we calculate for 3C 239 is consistent with local star-forming galaxies \citep[sSFR\,$\approx$\,0.25\,Gyr$^{-1}$,][]{Elbaz:11}, however, most of the other sources have much higher sSFRs and so they are building their mass at a faster rate. Assuming a constant SFR, most of the radio galaxies have taken $<$\,1\,Gyr to form their stars and are likely to have already formed the bulk of their mass. This is because galaxies with masses of $\sim$\,10$^{12}$\,M$_{\rm \odot}$ reside in the tail of the stellar mass function up to at least $z$\,=\,4, as shown in \citet{Perez-Gonzalez:08}. If the star formation in our sources continues at the rate we observe, it would take them $\sim$4\,Gyrs to reach the stellar mass of the giant elliptical galaxy M87 \citep[$\sim$\,10$^{12.5}$\,M$_{\rm \odot}$,][]{Wu:06}. If we assume the mass of M87 as an approximate limit to a galaxy's stellar mass, it is unlikely that our radio galaxies will maintain their current SFRs to the present day. A possible scenario for quenching the star formation is the evolutionary sequence proposed by \citet{Sanders:88}. This sequence is as follows: (i) through a major-merger event, there is a phase of extreme star formation, (ii) once a supply of gas has reached the central region, an obscured AGN arises, (iii) as the SFR declines, the AGN starts to dominate, it sheds its obscuring dust and a quasi-stellar object (QSO) is formed, (iv) once the QSO runs out of fuel, this culminates into a quiescence elliptical galaxy. The first stage of the \citet{Sanders:88} sequence is consistent with the proposition that powerful radio-loud AGN are a consequence of gas-rich mergers \citep{Heckman:86, Hardcastle:07}. After the QSO phase, it is now thought that the AGN inhibits further star formation through `radio-mode' feedback, in which the relativistic jets inject enough energy into the intergalactic medium to suppress the cooling of gas onto the AGN host galaxy and the eventual production of stars \citep{Bower:06, Croton:06}. The inclusion of SMBH activity in galaxy formation models therefore helps to explain why most local massive elliptical galaxies predominantly contain old and red stars. In the context of the \citet{Sanders:88} evolutionary sequence, the properties we measure for our sample of radio galaxies (high AGN luminosities and SFRs) could be the result of a merging event, however we now observe them at a point early on in the third stage of the \citet{Sanders:88} evolutionary sequence, where the obscured AGN dominates the system but star formation is still significant. The radio galaxy 3C 239 (and perhaps NVSS J180556$+$633313), with a relatively high stellar mass and low SFR, appears to be further along this evolutionary track than the rest of the sample, where a QSO is closer to forming. Furthermore, the short timescales suggested for the radio-loud phase of AGN \citep[$<$\,10$^{7}$\,years,][]{Blundell:99} in combination with the high stellar masses of our HzRGs, implies that they must have been forming their stars long before the onset of the radio emission. This suggests that the powerful radio-loud phase occurs near the end of the formation of the host galaxy, perhaps during the third stage of evolution proposed by \citet{Sanders:88}, but before the point at which host galaxy activity ceases completely, allowing the two phases to be observed simultaneously.

\section{Conclusion}
We have separated the mid-IR spectra of seven powerful high-redshift radio galaxies ($1.5\,<\,z\,<\,2.6$) into AGN and host galaxy components, observing PAH emission and silicate absorption for several sources. These HzRGs are powerful systems with high mid-IR AGN luminosities ($\nu L_{\rm \nu}$\,(6\,$\mu$m)\,$\approx$\,10$^{46}$\,erg\,s$^{-1}$ for the mean spectrum).

We find a wide range of SFRs, ranging from $\sim$\,1000\,M$_{\rm \odot}$\,yr$^{-1}$ to no significantly detected star formation and we constrain the mean SFR for the overall HzRG population in our redshift range (1.5\,$<$\,$z$\,$<$\,2.6) to be 340-790\,M$_{\rm \odot}$\,yr$^{-1}$. Thus, at the epoch we probe, some HzRGs are forming stars at a high rate. Comparing with the results from other studies, we observe that radio-loud AGN over a few orders of magnitude in radio luminosity have comparable SFRs. Overall we find no evidence for a significant correlation between radio emission and star formation in powerful radio-loud AGN at high redshift.

Attributing the silicate absorption feature entirely to the AGN torus, the relatively small absorption features ($\tau_{\rm 9.7\,\mu m}\,<$\,0.8) we derive for these objects is an indication that their tori have a clumpy distribution and contain relatively small number of clouds, $\sim$\,5-15, through the equatorial plane. These optical depths agree with those for some radio-quiet type-2 AGN and lower-redshift radio galaxies, implying the tori have similar properties. However, the greater silicate absorption features observed for radio-loud AGN-dominated ULIRGs and local Compton-thick AGN suggests the host galaxy can act as a cold dust screen for some sources.

The sSFRs we calculate range from $\sim$\,0.3 to 3\,Gyr$^{-1}$, although lower limits of $>$\,1\,Gyr$^{-1}$ are calculated for most of the sources in the sample. Assuming a constant SFR, this suggests that most of the sources have taken $<$\,1\,Gyr to form their stars. They are also likely to have already formed the bulk of their stellar mass and so we expect their SFRs to decline in the future.

To consider how these radio galaxies formed and evolved, they likely represent a stage where the obscured AGN has developed enough to dominate the system but the declining star formation is still significant.

Since the timescale of the radio emission is short and these radio galaxies are all massive, they must have started to form their stars a long time before the radio-loud phase commenced. This is an indication that the powerful radio-loud phase occurs near the end of the host galaxy activity but during a stage where significant star formation is still taking place.

\section{Acknowledgments}
We thank the anonymous referee and R.C. Hickox for their useful comments which improved the manuscript. JIR acknowledges the support of a Science and Technologies Facilities Council studentship. NS is the recipient of an Australian Research Council Future Fellowship. This work is based on observations made with the \textit{Spitzer Space Telescope}, which is operated by the Jet Propulsion Laboratory (JPL), California Institute of Technology (Caltech) under contract with NASA. Support for this work was provided by NASA through an award issued by JPL/Caltech. The IRS was a collaborative venture between Cornell University and Ball Aerospace Corporation funded by NASA through JPL and Ames Research Center. This work benefitted from the NASA/IPAC Extragalactic Database (NED), which is operated by the JPL, Caltech, under contract with NASA.

\bibliographystyle{mnras}
\bibliography{PAH_emission_in_HzRGs_references}
\label{lastpage}

\begin{landscape}
\begin{table}
\begin{center}{
\caption{The properties of the seven HzRGs in the sample. Columns are as follows: (1) HzRG name; (2) name as given by the SHzRG project; (3) and (4) J2000 co-ordinates; (5) spectroscopic redshift; (6) rest-frame 500\,MHz luminosity, with assumed uncertainty of 20\% ; (7), (8) and (9) observed 16\,$\mu$m, 24\,$\mu$m and 850\,$\mu$m flux densities, respectively; (10) stellar mass; (11) angular size of radio source, with assumed uncertainty of 10\%; (12) radio core dominance, R, defined as the ratio of core and extended radio flux at rest-frame 20\,GHz, with assumed uncertainty of 50\%.}
\begin {spacing} {1.5}
\begin{tabular}{@{}llcccccccccc@{}}
\hline
\hline
HzRG & SHzRG name$^{a}$ & RA$^{a,*}$ & Dec.$^{a,*}$ & $z$$^{a}$ & log(L$_{\rm 500\,MHz}$)$^{a}$ & S$_{\rm 16\,\mu m}$$^{a}$ & S$_{\rm 24\,\mu m}$$^{a}$ & S$_{\rm 850\,\mu m}$ & log(M$_{\rm *}$/M$_{\rm \odot}$)$^{a}$ & $\theta$$^{a}$ & R$^{b}$\\
& & [J2000] & [J2000] &  & [W\,Hz$^{-1}$] & [mJy] & [mJy] & [mJy] & & [arcsec]\\
(1) & (2) & (3) & (4) & (5) & (6) & (7) & (8) & (9) & (10) & (11) & (12)\\
\hline
3C 239 & 3C 239 & 10:11:45.42 & +46:28:19.8 & 1.781 & 29.00 & 0.85\,$\pm$\,0.09 & 1.89\,$\pm$\,0.06 & 0.00\,$\pm$\,1.00$^{c,\dag}$ & 11.60 & 11.9 & 0.003\\
4C 23.56 & 4C 23.56 & 21:07:14.80 & +23:31:45.0 & 2.483 & 28.93 & 2.40\,$\pm$\,0.09 & 4.63\,$\pm$\,0.04 & 1.68\,$\pm$\,0.98$^{c,\dag}$ & $<$\,11.59 & 53.0 & 0.171\\
NVSS J180556$+$633313 & 7C 1805$+$6332 & 18:05:56.81 & +63:33:13.1 & 1.840 & 27.78 & 0.31\,$\pm$\,0.08 & 0.67\,$\pm$\,0.03 & $-$ & $<$\,11.07 & 15.6 & 0.006\\
NVSS J171414$+$501530 & LBDS 53W002 & 17:14:14.79 & +50:15:30.6 & 2.393 & 27.78 & 0.59\,$\pm$\,0.10 & 0.65\,$\pm$\,0.04 & 0.99\,$\pm$\,1.10$^{c,\dag}$ & $<$\,11.27 & 1.5 & $-$\\
PMN J0408$-$2418 & MRC 0406$-$244 & 04:08:51.46 & $-$24:18:16.4 & 2.427 & 29.03 & 0.64\,$\pm$\,0.09 & 1.54\,$\pm$\,0.04 & $-$ & 11.38 & 10.0 & 0.027\\
PKS 1138$-$26 & PKS 1138$-$262 & 11:40:48.38 & $-$26:29:08.8 & 2.156 & 29.07 & 3.02\,$\pm$\,0.10 & 3.89\,$\pm$\,0.02 & 6.7\,$\pm$\,2.4$^{d,\ddag}$ & $<$\,12.26 & 15.8 & 0.032\\
4C $-$00.54 & USS 1410$-$001 & 14:13:15.10 & $-$00:22:59.7 & 2.363 & 28.41 & 0.66\,$\pm$\,0.10 & 1.07\,$\pm$\,0.04 & $-$ & $<$\,11.41 & 24.0 & 0.072\\
\hline
\end{tabular}

\end{spacing}
\label{Table:source_info}}
\end{center}
$^{*}$J2000 co-ordinates from IRAC observations at 3.6\,$\mu$m.\\
$^{\dag}$850\,$\mu$m flux densities have been corrected for radio synchrotron emission contamination.\\
$^{\ddag}$The 850\,$\mu$m flux density of 12.8\,$\pm$\,3.3\,mJy for PKS 1138$-$26 from \citet{Reuland:04} is the result of a poor quality observation.
\begin {spacing} {1.3}
\textbf{References.} --- (a) DB10 and references therein; (b) Drouart et al. (2012; \textit{in prep.}) and references therein; (c) \citet{Archibald:01}; (d) \citet{Stevens:03}.
\end{spacing}
\end{table}

\begin{table}
\begin{center}{
\caption{The first and second order observations with the IRS Long-Low (LL) module for the sample.}
\begin{tabular}{@{}lcc@{}}
\hline
\hline
HzRG & LL First Order & LL Second Order\\
& [seconds] & [seconds]\\
\hline
3C 239$^{a}$ & 10 x 2 x 120 & 10 x 2 x 120\\
4C 23.56$^{b}$ & 12 x 2 x 120 & $-$\\
NVSS J180556$+$633313$^{a}$ & 20 x 2 x 120 & 20 x 2 x 120\\
NVSS J171414$+$501530$^{a}$ & 20 x 2 x 120 & $-$\\
PMN J0408$-$2418$^{c}$ & 15 x 2 x 120 & 15 x 2 x 120\\
PKS 1138$-$26$^{d}$ & 13 x 8 x 120 & $-$\\ 
4C $-$00.54$^{a}$ & 15 x 2 x 120 & $-$\\
\hline
\end{tabular}
\label{Table:source_obs}}
\end{center}
\textbf{Observations.} --- (a) N. Seymour (PID 50379); (b) G. Fazio (PID 40093); (c) N. Nesvadba (PID 50402); (d) P. Ogle (PID 40460).
\end{table}
\end{landscape}

\begin{landscape}
\begin{table}
\begin{center}{
\caption{Results from the spectral fitting process with 1\,$\sigma$ uncertainties. Columns are as follows: (1) HzRG name; (2) 7.7\,$\mu$m PAH luminosity; (3) host galaxy component contribution to the best fit in the 24\,$\mu$m MIPS band; (4) total IR (8-1000\,$\mu$m) star-forming luminosity, inferred from 7.7\,$\mu$m PAH luminosity; (5) star formation rate, inferred from 7.7\,$\mu$m PAH luminosity; (6) monochromatic rest-frame 6\,$\mu$m AGN luminosity, (7) mid-IR spectral index of AGN component, (8) 9.7\,$\mu$m AGN optical depth, (9) specific star formation rate; (10) best-fitting \protect \citet{Brandl:06} template; (11) $\chi^{2}$/degrees of freedom, $\nu$.}
\begin {spacing} {2.0}
\tabcolsep 5.5pt
\begin{tabular}{@{}lcccccccccc@{}}
\hline
\hline
HzRG & $L_{\rm 7.7 PAH}$ & Host galaxy & $L_{\rm IR}$/L$_{\rm \odot}$$^{a}$ & SFR & $\nu L_{\rm \nu}$\,(6\,$\mu$m) & $\alpha$ & $\tau_{\rm 9.7\,\mu m}$ & sSFR & Best-fit host & $\chi^{2}$\,/\,$\nu$\\
& & contribution & & & & & & & galaxy template &\\
& [10$^{10}$\,L$_{\rm \odot}$] & [\%] & [10$^{12}$\,L$_{\rm \odot}$] & [M$_{\rm \odot}$\,yr$^{-1}$] & [10$^{45}$\,erg\,s$^{-1}$] & & & [Gyr$^{-1}$]\\
(1) & (2) & (3) & (4) & (5) & (6) & (7) & (8) & (9) & (10) & (11)\\
\hline
Mean spectrum              & 2.69\,$\pm$\,0.15          & $-$                        & 3.61\,$^{+\,3.62}_{-\,1.82}$     & 620\,$^{+\,620}_{-\,310}$     & 8.97\,$^{+\,0.17}_{-\,0.14}$    & 1.00\,$^{+\,0.04}_{-\,0.00}$   & 0.01\,$^{+\,0.02}_{-\,0.01}$   & $-$                                   & NGC2623   & 162\,/\,48\\
3C 239                           & 0.84\,$^{+\,0.22}_{-\,0.19}$  & 13\,$\pm$\,3        & 1.06\,$^{+\,1.10}_{-\,0.59}$     & 180\,$^{+\,190}_{-\,100}$     & 4.45\,$\pm$\,0.17            & 1.50\,$^{+\,0.00}_{-\,0.02}$   & 0.33\,$\pm$\,0.04           & 0.31\,$^{+\,0.32}_{-\,0.17}$  & Mrk52         & 270\,/\,144\\
4C 23.56                        & 4.23\,$^{+\,0.63}_{-\,0.69}$  & 8\,$\pm$\,1          & 5.82\,$^{+\,5.88}_{-\,3.06}$     & 1000\,$^{+\,1010}_{-\,530}$  & 28.17\,$^{+\,0.53}_{-\,0.48}$  & 1.17\,$\pm$\,0.11           & 0.12\,$\pm$\,0.06           & $>$\,1.72                        & NGC2623    & 221\,/\,81\\
NVSS J180556$+$633313  & 0.13\,$^{+\,0.21}_{-\,0.13}$  & 5\,$^{+\,8}_{-\,5}$     & 0.15\,$^{+\,0.28}_{-\,0.15}$     & 20\,$^{+\,50}_{-\,20}$          & 1.65\,$\pm$\,0.12            & 1.11\,$\pm$\,0.12           & 0.00\,$^{+\,0.02}_{-\,0.00}$   & $-$                                  & NGC4088    & 122\,/\,141\\
NVSS J171414$+$501530  & 1.37\,$^{+\,1.17}_{-\,0.28}$  & 23\,$^{+\,20}_{-\,5}$  & 1.77\,$^{+\,2.33}_{-\,0.96}$    & 310\,$^{+\,400}_{-\,160}$     & 2.69\,$^{+\,0.19}_{-\,0.69}$     & 1.50\,$^{+\,0.00}_{-\,0.07}$   & 0.00\,$^{+\,0.08}_{-\,0.00}$   & $>$\,1.09                        & NGC2623    & 74\,/\,81\\
PMN J0408$-$2418            & 4.71\,$^{+\,0.48}_{-\,0.40}$  & 36\,$^{+\,4}_{-\,3}$   & 6.52\,$^{+\,6.54}_{-\,3.30}$    & 1120\,$^{+\,1130}_{-\,570}$  & 5.76\,$^{+\,0.40}_{-\,0.43}$     & 1.50\,$^{+\,0.00}_{-\,0.30}$   & 0.70\,$^{+\,0.09}_{-\,0.17}$   & 3.12\,$^{+\,3.14}_{-\,1.58}$  & NGC2623    & 123\,/\,103\\
PKS 1138$-$26                  & 4.80\,$^{+\,0.41}_{-\,0.37}$   & 18\,$^{+\,2}_{-\,1}$   & 6.65\,$^{+\,6.67}_{-\,3.36}$    & 1150\,$^{+\,1150}_{-\,580}$  & 14.45\,$^{+\,0.54}_{-\,0.48}$   & 1.41\,$\pm$\,0.07           & 0.00\,$^{+\,0.02}_{-\,0.00}$   & $>$\,0.42                        & NGC2623    & 129\,/\,81\\
4C $-$00.54                      & 2.02\,$^{+\,0.67}_{-\,0.61}$   & 21\,$\pm$\,7        & 2.67\,$^{+\,2.81}_{-\,1.56}$    & 460\,$^{+\,480}_{-\,270}$      & 4.65\,$^{+\,0.61}_{-\,0.57}$     & 1.00\,$^{+\,0.11}_{-\,0.00}$   & 0.08\,$^{+\,0.11}_{-\,0.08}$   & $>$\,1.19                        & NGC4088    & 107\,/\,81\\
\hline
\end{tabular}
\end{spacing}
\label{Table:fitting_results}}
\end{center}
$^{a}$We take into account the 1\,$\sigma$ scatter in the $L_{\rm 7.7 PAH}-L_{\rm IR}$ relation.
\begin {spacing} {1.6}
\textbf{Note.} --- A calibration uncertainty of $\sim$\,5\,\% is not included for the measurements in columns 2, 4, 5, 6 and 9.
\end{spacing}
\end{table}
\end{landscape}

\end{document}